\documentclass[prd,preprintnumbers,nofootinbib,tightenlines,superscriptaddress]{revtex4-2}
\usepackage{graphicx,epstopdf}
\usepackage{amsmath,amssymb}
\usepackage[usenames, dvipsnames]{color}
\usepackage{slashed}
\usepackage{times}
\usepackage{latexsym}
\usepackage{longtable}
\usepackage[utf8]{inputenc}
\usepackage{hyperref}
\DeclareMathOperator{\arctanh}{arctanh}
\newcommand{\orcid}[1]{\href{https://orcid.org/#1}{\includegraphics[width=8pt]{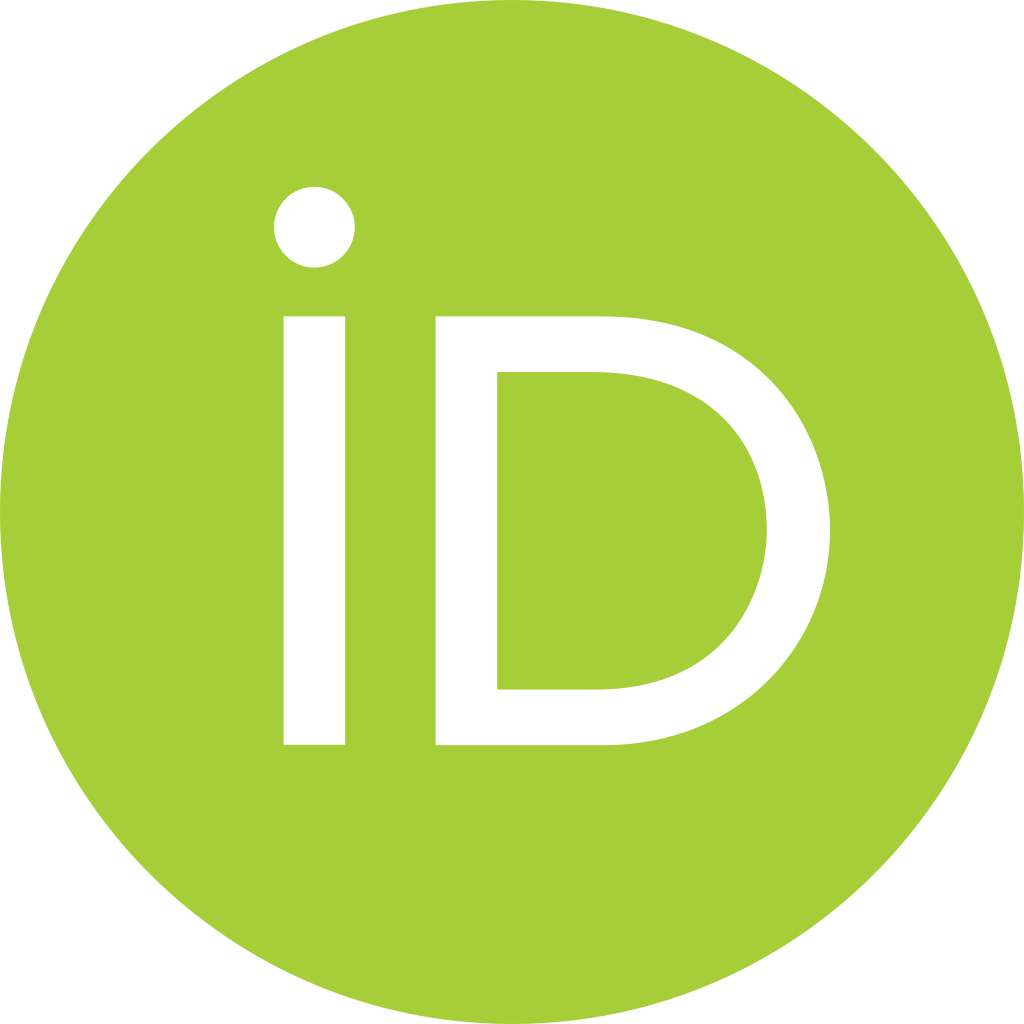}}}
\begin{document}


\title{Probing Superheavy Dark Matter with Exoplanets}

\author{Mehrdad Phoroutan-Mehr \orcid{0000-0002-9561-0965}}
\email{mphor001@ucr.edu}
\affiliation{Department of Physics and Astronomy, University of California, Riverside, CA 92521, USA}

\author{Tara Fetherolf \orcid{0000-0002-3551-279X}}
\email{tara.fetherolf@ucr.edu}
\affiliation{Department of Earth and Planetary Sciences, University of California, Riverside, CA 92521, USA}
\affiliation{NASA Postdoctoral Program Fellow}


\date{\today}


\begin{abstract}
Exoplanets, with their large volumes and low temperatures, are ideal celestial detectors for probing dark matter (DM) interactions. DM particles can lose energy through scattering with the planetary interior and become gravitationally captured if their interaction with the visible sector is sufficiently strong. In the absence of annihilation, the captured DM thermalizes and accumulates at the planet's center, eventually collapsing into black holes (BHs). Using gaseous exoplanets as an example, we demonstrate that BH formation can occur within an observable timescale for superheavy DM with masses greater than $10^6 \, {\rm GeV}$ and nuclear scattering cross sections. The BHs may either accrete the planetary medium or evaporate via Hawking radiation, depending on the mass of the DM that formed them. We explore the possibility of periodic BH formation within the unconstrained DM parameter space and discuss potential detection methods, including observations of planetary-mass objects, pulsed high-energy cosmic rays, and variations in exoplanet temperatures. Our findings suggest that future extensive exoplanet observations could provide complementary opportunities to terrestrial and cosmological searches for superheavy DM.
\end{abstract}


\maketitle


\section{Introduction}
Detecting Dark Matter (DM) via non-gravitational interactions remains a central focus of ongoing searches for new physics beyond the Standard Model (SM). Terrestrial experiments and astrophysical observations have imposed stringent constraints on the parameter space of the prevailing Weakly Interacting Massive Particle (WIMP) model. While extensive efforts have been dedicated to theoretical models and experimental searches for light DM \cite{Knapen:2017xzo, Zurek:2024qfm, Essig:2022dfa}, DM candidates heavier than the weak scale present another compelling direction within the DM parameter space for future searches \cite{Carney:2022gse}. The traditional upper bound on the mass of thermal DM is $\mathcal{O}(100) \, {\rm GeV}$ set by the perturbative unitarity in their production processes \cite{Griest:1989wd}. Recent studies have proposed alternative cosmological mechanisms for the production of DM with masses beyond the unitarity bound \cite{Berlin:2016gtr, Kim:2019udq, Berlin:2017ife, Kramer:2020sbb,Frumkin:2022ror, Chway:2019kft, Baker:2019ndr, Asadi:2021yml, Lennon:2017tqq, Morrison:2018xla, Hooper:2019gtx, Gehrman:2023qjn}. Experimental searches for these superheavy DM candidates, spanning from $100 \, {\rm TeV}$ to the Planck scale $M_{\rm pl}$, remain challenging due to the suppressed DM flux in direct detection experiments, necessitating an extremely large exposure that is beyond the reach of current ground-based detectors. This has motivated the exploration of superheavy DM using astrophysical objects as targets through scattering and capture processes, driven by the same DM-SM interactions probed in terrestrial experiments. To date, extensive studies have explored the capture of DM particles by various astrophysical bodies, leading to a range of consequences, including heating, cooling, Black Hole (BH) formation in the core, neutrino emission, alterations in the equation of state of compact objects and their mass-radius relation, as well as gravitational wave emission. These astrophysical entities include neutron stars\cite{Goldman:1989nd, Gould:1989gw, Kouvaris:2007ay, McDermott:2011jp, Bell:2013xk, Bramante:2017ulk, Baryakhtar:2017dbj, Raj:2017wrv, Bell:2018pkk, Acevedo:2019agu, Joglekar:2019vzy, Bell:2019pyc, Bell:2020obw, Bell:2020lmm, Bell:2020jou, Joglekar:2020liw, Anzuini:2021lnv, Abac:2021txj, Leane:2021ihh, Nguyen:2022zwb, Alvarez:2023fjj, Bell:2023ysh, Bhattacharya:2023stq,Linden:2024uph, Acevedo:2024ttq, Basumatary:2024uwo, Lu:2024kiz, Liu:2024qbe}, white dwarfs\cite{McCullough:2010ai, Graham:2018efk, Acevedo:2019gre, Dasgupta:2019juq, Bell:2021fye, Ramirez-Quezada:2022uou, Acevedo:2023xnu, Bhattacharya:2024pmp, Niu:2024nws, Bell:2024qmj}, the Sun\cite{Bottino:2002pd, Bernal:2012qh, Chen:2015uha, Feng:2016ijc, Lucente:2017jcp, Arina:2017sng, Leane:2017vag, Niblaeus:2019gjk, Cuoco:2019mlb, Nunez-Castineyra:2019odi, Mazziotta:2020foa, Bell:2021pyy, DUNE:2021gbm, Ray:2023auh,Chu:2024gpe, Nguyen:2025ygc}, the Earth\cite{Mack:2007xj, Chauhan:2016joa, Acevedo:2020gro}, the Moon\cite{Garani:2019rcb, Chan:2020vsr}, and other planetary bodies\cite{Kawasaki:1991eu, Mitra:2004fh, Adler:2008ky, Bramante:2019fhi, Bramante:2022pmn, Blanco:2023qgi, Blanco:2024lqw, Benito:2024yki, Acevedo:2024zkg,Blanco:2024lqw,Robles:2024tdh}.

Celestial bodies accelerate DM particles towards them via gravitational forces, potentially causing the DM particles to scatter off the substances composing these bodies. If the scattered DM particles lose sufficient kinetic energy, they become gravitationally bound to these compact objects. In annihilating DM models, the captured DM can annihilate into Standard Model particles, injecting heat into these celestial objects and emitting weakly interacting particles such as neutrinos. Assuming the annihilation or decay of captured DM is negligible (e.g. if the relic DM is asymmetric, or its stability is protected by symmetry), the accumulation of DM due to capture can continue indefinitely. The captured DM particles thermalize through ongoing multiple scattering events and, depending on the particle model of the DM, can drift to the core of these objects, forming a clump of DM. In such cases, the accumulated DM particles can surpass the critical mass and form a BH at the center. Recently, the formation of BHs in the cores of Earth and the Sun has been studied in \cite{Acevedo:2020gro} as well as other celestial objects in \cite{Ray:2023auh} in the context of superheavy DM models. The evolution of the formed BH can have significant consequences. For instance, if the BH grows, it can eventually destroy the compact object. Conversely, if the size of the formed BH is small enough, it can evaporate due to Hawking radiation. This evaporation process can lead to observable consequences and, depending on the DM parameter space, may result in neutrino emissions, high-energy photon and electron-positron emissions, or heating of the host body.

Exoplanets, similar to other compact astrophysical bodies, sweep up DM particles along their path as they move through the galactic halo, potentially capturing DM particles. Exoplanets are celestial bodies primarily formed from the protoplanetary disk of gas and dust surrounding a young star. Giant exoplanets, which reach the gas accretion phase, are primarily composed of hydrogen and helium, and nuclear fusion does not occur within their interior. Exoplanets can either exist as bound companions to a host star within an exosolar system or as rogue planets (also known as free-floating planets) drifting through interstellar space. Exoplanets are promising candidates for DM capture studies for several reasons. Firstly, exoplanets have a larger radius compared to other compact object targets, allowing for a higher DM interaction rate through their geometric cross section. For example, Jupiter has a mass of $M_\mathrm{J} = 1.9 \times 10^{30} \, {\rm g}$ and a radius of $R_{\rm J} = 7 \times 10^4 \, {\rm km}$, making it four orders of magnitude larger in size than neutron stars. The larger surface area also supports the detection of distant exoplanets via their surface emissions. Secondly, exoplanets are abundant and distributed throughout the entire galaxy. Thus far, over $5000$ exoplanets have already been discovered \cite{2013PASP..125..989A}, with many more anticipated in the near future, given the estimate that, on average, each star is accompanied by 1.6 exoplanets in the Milky Way \cite{2012Natur.481..167C}. Exoplanets observed at locations closer to the Galactic Center (GC) are particularly interesting since they are exposed to a higher DM number density, showing significant potential to set stronger constraints on small DM interaction strength than targets in the Solar neighborhood. The SWEEPS and MOA missions have discovered exoplanets in the galactic bulge \cite{Sahu:2006ex, 2010ApJ...711..731J,2023AJ....166..108S}. Detection of exoplanets towards the GC direction is one of the goals of Nancy Grace Roman Space Telescope \cite{Spergel:2015sza} with its Galactic Bulge Time Domain Survey. 
The nearby searches will also be extended by the future Habitable Worlds Observatory \cite{nasaHWO} to include a $1 \, {\rm kpc}$ local volume. Moreover, isolated exoplanets and those far from their host stars remain cold, making their surface temperature an ideal observable for detecting energy injection from DM. This contrasts with heavier astrophysical objects, which are heated by nuclear reactions. Indeed, exoplanets can theoretically approach extremely low surface temperatures after sufficient cooling over time, allowing any additional heat source to be identified as an anomalous surface temperature. Overheated exoplanets have recently been studied to constrain light DM models with infrared telescope observations \cite{Leane:2020wob}. The heat from DM annihilation can also modify the formation of planets such that their existence sets constraints on DM~\cite{Croon:2023bmu}. Additionally, the captured DM abundance is studied with their contributions to the heat transport rate of brown dwarfs in \cite{Banks:2024eag}.

In this work, we revisit the impact of non-annihilating superheavy DM on gaseous exoplanets, with a particular focus on the capture of DM particles by these exoplanets. We focus on the formation of BHs within exoplanets and examine the conditions under which a BH can form, grow, or evaporate. In the absence of depletion in the accumulated DM, the BH formation rate is determined by the DM capture rate and its evolution, including thermalization, drifting, and collapsing timescales. Specifically, BH formation occurs more rapidly for heavy DM candidates, even though the capture rate is smaller. Therefore, in this work, we focus on superheavy, non-annihilating DM models. Furthermore, we introduce a simple model for exoplanets and investigate the sensitivity of the bounds to various assumptions made in modeling them. The formation of a BH depends on the particle model of DM as well as the characteristics of exoplanets. In this regard, we have assumed both fermionic and bosonic DM models, with and without coherent effects in the scattering cross section off SM particles. Additionally, we have considered exoplanets of varying masses at different distances from the GC, along with variations in the temperature and density profiles in their modeling. The adopted exoplanet mass ranges from Jupiter-sized to those $13$ times larger, and we extend this to $30$ times larger. We show that these bounds are complementary to other studies, and in some regions, they are stronger. Furthermore, we identify the DM parameter space where the formed BH can grow and destroy exoplanets, remain stable, or evaporate. Heavier exoplanets closer to the center of the galaxy possess a greater capability to form a BH within them. Additionally, we demonstrate that, in certain regions of the parameter space that are still allowed by current bounds, a BH could form in as little as $10$ months.

This paper is organized as follows. In Sec.~\ref{S:Exoplanets}, we describe the model for the structure of gaseous exoplanets and their observation methods. In Sec.~\ref{S:CaptureRate}, we outline the calculation of DM capture by exoplanets. In Sec.~\ref{S:BHFormationCondition}, we examine the conditions necessary for the captured DM to collapse into a BH. Finally, in Sec.~\ref{S:Results}, we present the DM mass and scattering cross section parameter space where a BH can form within an exoplanet. The conclusions are summarized in Sec.~\ref{S:Conclusion}.


\section{Gaseous Planet Model and Observations}
\label{S:Exoplanets}


\subsection{Gas Planet Model}
\label{S:ExoplanetStructure}
In this section, we introduce the exoplanet model adopted in this work. The modeling of planets is an actively advancing field, with extensive literature studying their properties, including internal structure, chemical composition, and equations of state (we refer the readers to \cite{MARLEY2014743} and references therein). The structure of Jupiter in the Solar System has been discussed extensively in various studies~\cite{2019ApJ...872..100D, MARLEY2014743, 2022A&A...664A.112M, kerley2013structures, 2012ApJS..202....5F, Mitra2018, WEST2014723}. Additional theoretical work has explored the internal structure of objects across a wide range of planetary masses, as shown in \cite{1995ApJS...99..713S, 2022arXiv221106518Z, 2024arXiv240104172T, 2013ApJ...774..148M, Auddy_2016, 2023A&A...671A.119C, 2019ApJ...872...51C, 2021ApJ...917....4C}. Nevertheless, general knowledge of the density and temperature profiles of exoplanets remains limited.

In this work, we follow \cite{2019ApJ...872...51C, 2021ApJ...917....4C} to model the equation of state of gaseous planet in order to obtain their internal structure profiles. This reference presents a novel equation of state for dense hydrogen and helium mixtures, spanning the temperature-density domain from solar-size stars to brown dwarfs and gaseous planets. The analysis in \cite{2019ApJ...872...51C, 2021ApJ...917....4C} incorporates a more rigorous treatment for the mixture system of hydrogen and helium with the contribution from their interactions to the entropy of the system. This is particularly relevant for low-temperature, high-density systems characteristic of gas giant exoplanets. Furthermore, this analysis assumes no external heating, which is consistent with the exoplanet candidates used in this study. 
The density range covered in Ref.~\cite{2019ApJ...872...51C} and \cite{2021ApJ...917....4C} spans $10^{-6} \, \mathrm{g} \, \mathrm{cm^{-3}} \leq \rho \leq 10^{6} \, \mathrm{g} \, \mathrm{cm^{-3}}$, with temperatures ranging from $10^{2} \, \mathrm{K}\leq T \leq 10^{8} \, \mathrm{K}$ and pressures from $10^{-9} \, \mathrm{GPa} \leq P \leq 10^{13} \, \mathrm{GPa}$, and we obtained the data from~\cite{EOS}. The original data is computed for three different values of helium mass fraction $Y$, and we select $Y = 0.275$ that closely matches Jupiter's chemical composition. With this helium fraction, the average atomic mass is $\bar{m} = 1.18 \, \mathrm{GeV}$, the average atomic mass number is $\bar{A} = 1.26$, and the number of electrons per atom is $1.09$.

\begin{figure}[t!]
    \centering
    \includegraphics[width=0.32\columnwidth]{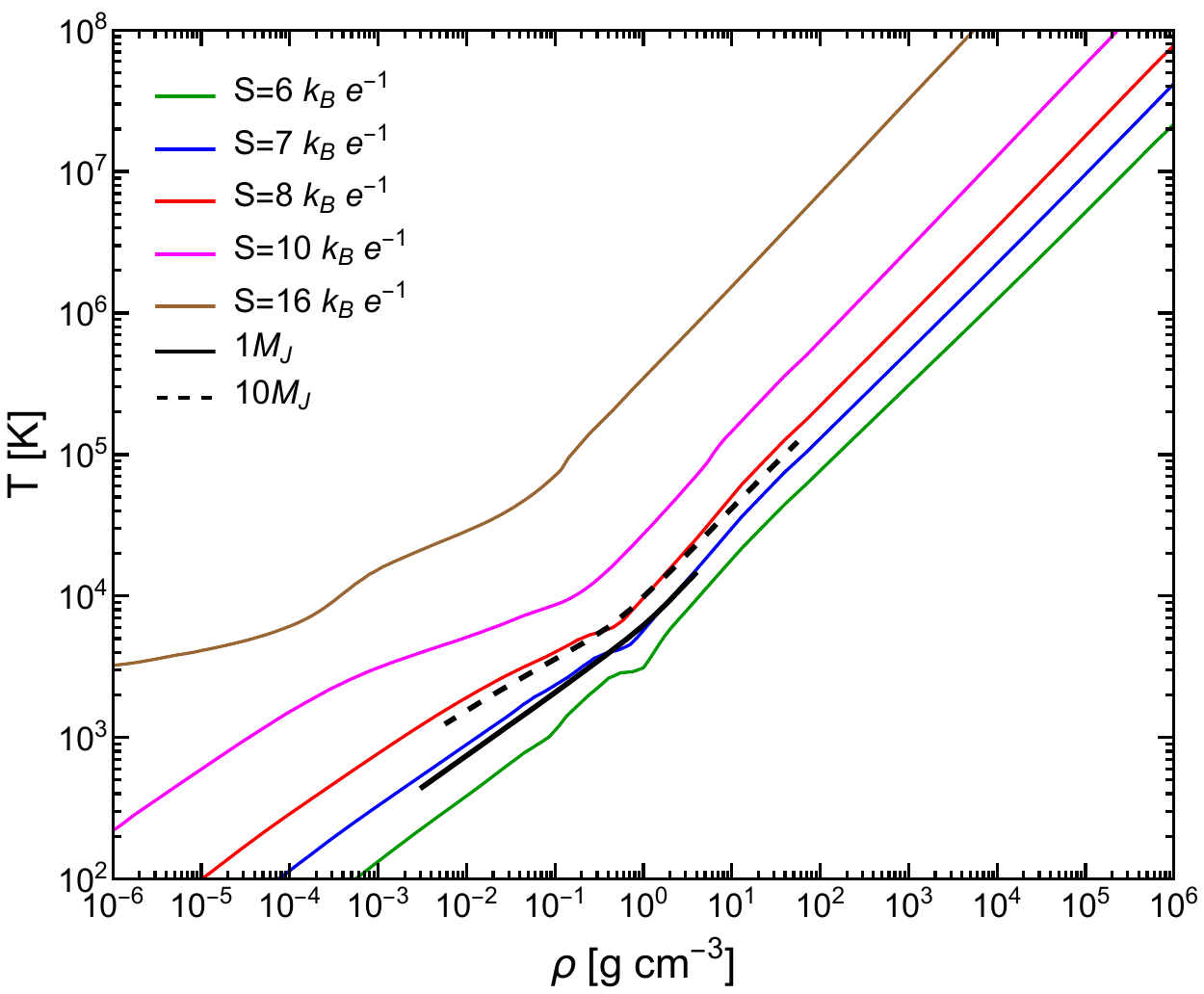} \,\,\,
    \includegraphics[width=0.32\columnwidth]{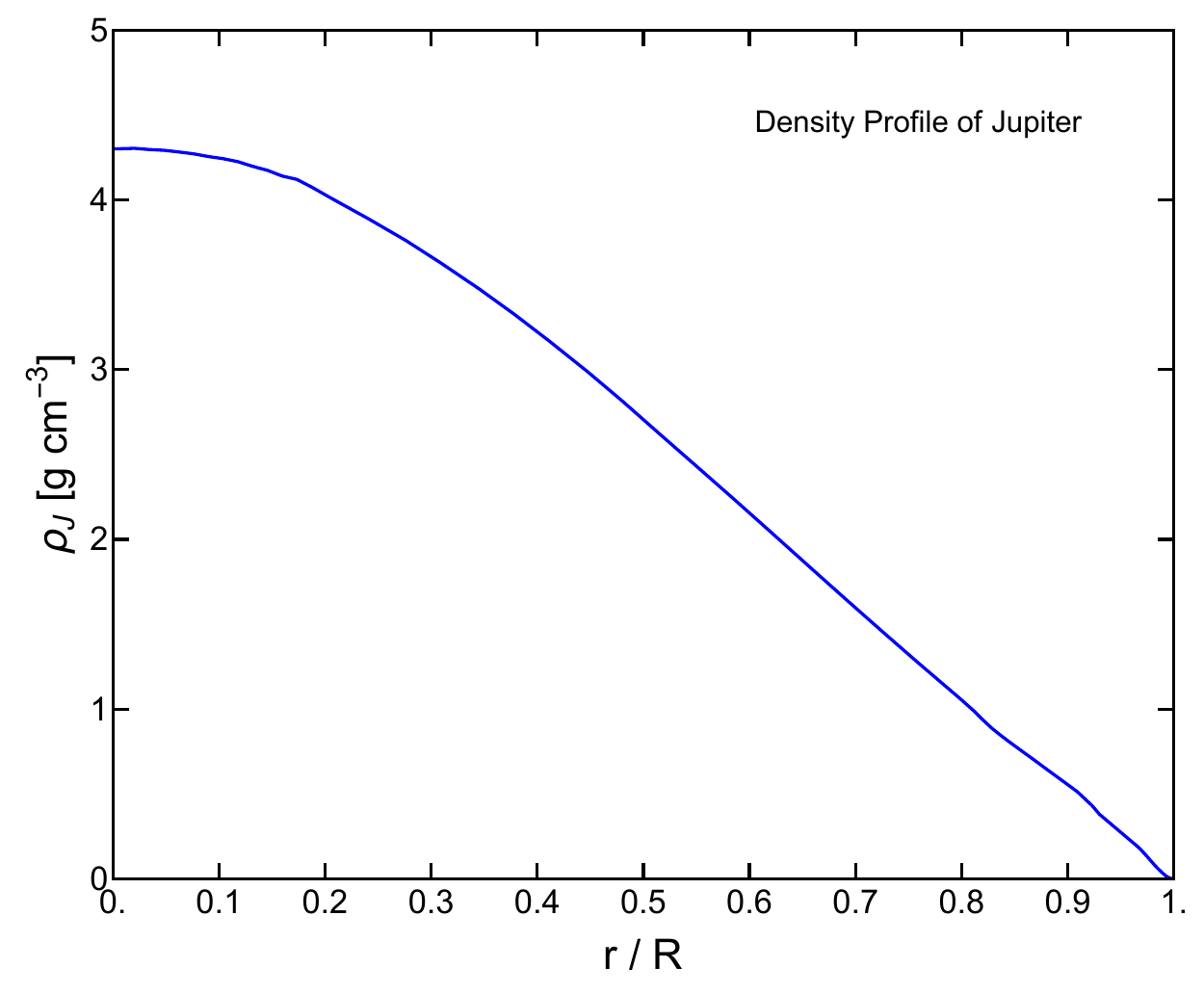} \,\,\,
    \includegraphics[width=0.32\columnwidth]{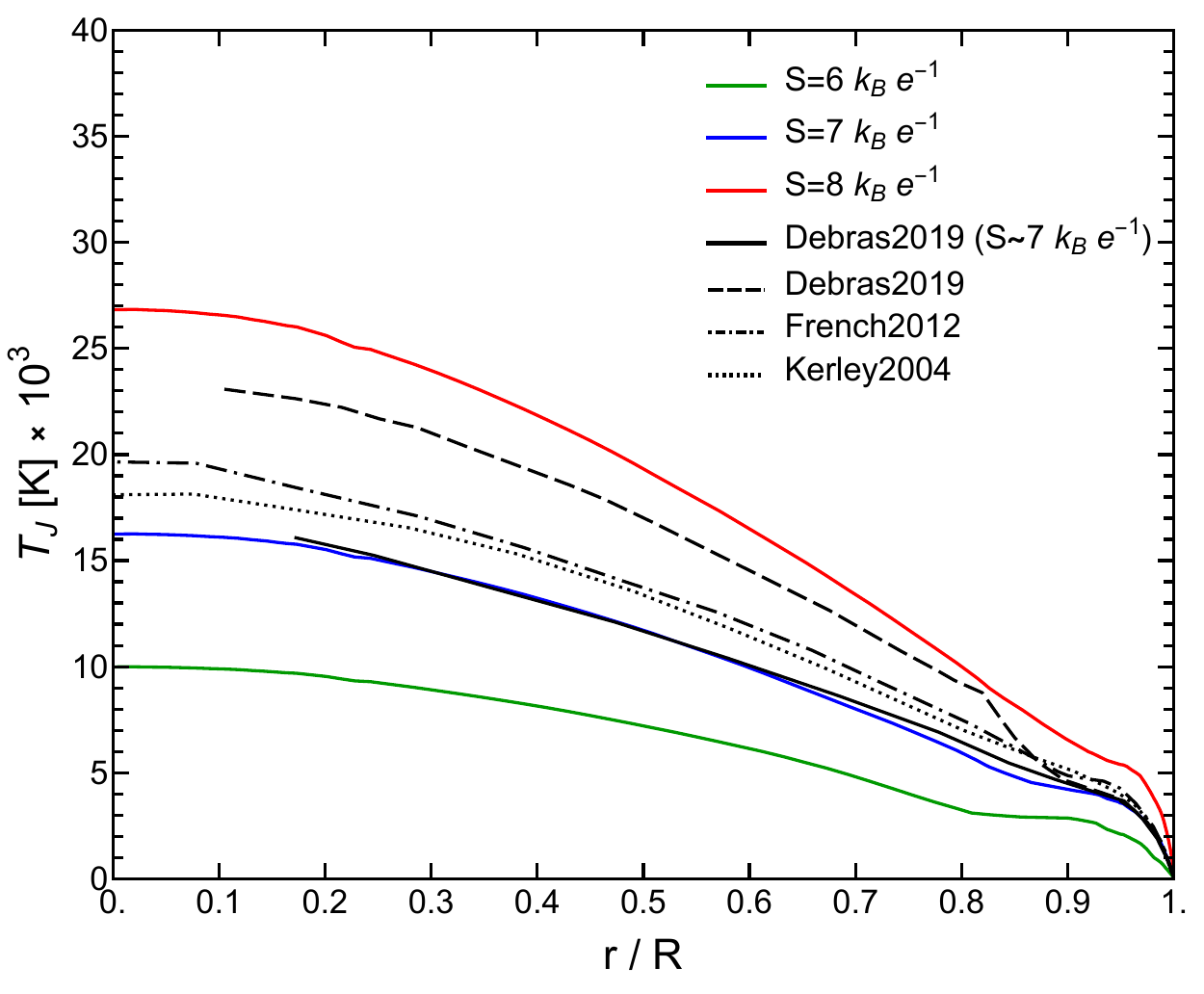}
    \caption{Left panel: Temperature-density relation of the hydrogen/helium mixture for constant entropy values of $S=6\, k_\mathrm{B} \, \mathrm{e^{-1}}$ (green), $7\, k_\mathrm{B} \, \mathrm{e^{-1}}$ (blue), $8\, k_\mathrm{B} \, \mathrm{e^{-1}}$ (red), $10\, k_\mathrm{B} \, \mathrm{e^{-1}}$ (magenta), and $16\, k_\mathrm{B} \, \mathrm{e^{-1}}$ (brown) in the isentropic model. Middle panel: Density profile of Jupiter without a core, adopted from \cite{MARLEY2014743}. Right: Temperature profiles of Jupiter-sized exoplanets assuming the entropy is fixed throughout the entire planet. For comparison, we plot the models from previous studies with a constant entropy model (solid black)~\cite{2019ApJ...872..100D}, as well as different varying-entropy models obtained from~\cite{2019ApJ...872..100D} (dashed black), ~\cite{2012ApJS..202....5F} (dot-dashed black), and~\cite{kerley2013structures} (dotted black).}
    \label{F:Density_Temperature_Jupiter}
\end{figure}

To simplify the analysis, we adopt an isentropic model, assuming constant entropy throughout the interiors of exoplanets, thereby avoiding complexities associated with entropy variations across different radii. We set the specific entropy per electron to $S=7\, k_\mathrm{B} \, \mathrm{e^{-1}}$  (or equivalently, $S \simeq 7.61 \, k_\mathrm{B} \, \mathrm{atom^{-1}} \simeq 0.05 \, \mathrm{MJ} \, \mathrm{kg^{-1}} \, \mathrm{K^{-1}}$), characteristic of Jupiter at its current age \cite{2019ApJ...872...51C, 2021ApJ...917....4C, 2019ApJ...872..100D}. Here $k_\mathrm{B}$ is the Boltzmann constant. More complex models of Jupiter, featuring varying entropy at different radii, have been studied in \cite{2019ApJ...872..100D, 2012ApJS..202....5F, kerley2013structures}. However, deviations from constant entropy do not significantly affect the final results of this work. We show the exoplanet model in Fig. \ref{F:Density_Temperature_Jupiter}. The left panel shows the temperature-density relationship of the hydrogen-helium mixture corresponding to constant entropy values. For comparison, we present different benchmark values: $6\, k_\mathrm{B} \, \mathrm{e^{-1}}$ (green), $7\, k_\mathrm{B} \, \mathrm{e^{-1}}$ (blue), $8\, k_\mathrm{B} \, \mathrm{e^{-1}}$ (red), $10\, k_\mathrm{B} \, \mathrm{e^{-1}}$ (magenta), and $16\, k_\mathrm{B} \, \mathrm{e^{-1}}$ (brown), adopted from the aforementioned data. Also shown are the temperature-density contour for two objects with masses equal to $1\,M_{\rm J}$ (solid black) and $10~M_{\rm J}$ (dashed black), as presented in \cite{2019ApJ...872...51C, 2021ApJ...917....4C}, for comparison with the isentropic model. As illustrated, the Jupiter-mass object closely aligns with the isentropic contour of $7 \, k_\mathrm{B} \, \mathrm{e^{-1}}$, while the $10 M_\mathrm{J}$ object aligns with the isentropic contour of $8 \, k_\mathrm{B} \, \mathrm{e^{-1}}$. 

To validate the model, we examine the density and temperature profiles with the known Jupiter’s properties. Jupiter’s density profile as a function of radius, derived from \cite{MARLEY2014743}, is shown in the middle panel of Fig.~\ref{F:Density_Temperature_Jupiter}. We follow the same method to truncate the surface density at $10^{-4} \, \mathrm{g} \, \mathrm{cm}^{-3}$ in order to avoid the zero surface temperature condition appeared in the isentropic model. The corresponding surface temperature, for $S = 7 \, k_\mathrm{B} \, {\rm e}^{-1}$, is about $114 \, \mathrm{K}$, being consistent with the observed surface temperature of Jupiter. In this profile model, the core density is about $4.3 \, \mathrm{g} \, \mathrm{cm}^{-3}$ and thus the central temperature is about $1.6 \times 10^{4} \, \mathrm{K}$, agreeing with Jupiter's core temperature. The right panel of Fig. \ref{F:Density_Temperature_Jupiter} depicts the temperature profile for Jupiter-like exoplanets, assuming the entropy is set to $S = 6 \, k_\mathrm{B} \, \mathrm{e}^{-1}$ (green),  $S = 7 \, k_\mathrm{B} \, \mathrm{e}^{-1}$ (blue), and  $S = 8 \, k_\mathrm{B} \, \mathrm{e}^{-1}$ (red). For comparison, we show the models presented in ~\cite{2019ApJ...872..100D} for constant entropy (solid black) and varying entropy (dashed black). Additionally, we include the models with varying entropy from \cite{2012ApJS..202....5F} (dot-dashed black) and \cite{kerley2013structures} (dotted black). As the plot illustrates, the isentropic model with $S = 7 \, k_\mathrm{B} \, \mathrm{e}^{-1}$ aligns well with the radial temperature profile in models referenced, among the available entropy value choices.

In this study, we focus on exoplanets within the mass range of $1 M_\mathrm{J}$ to $30 M_\mathrm{J}$. The so-far confirmed exoplanets listed in the NASA Exoplanet Archive \cite{2013PASP..125..989A} include those with masses up to $30 M_\mathrm{J}$. From the perspective of planetary formation and evolution, objects in the mass range between $13 M_\mathrm{J}$ and $30 M_\mathrm{J}$ are often considered brown dwarfs rather than gaseous exoplanets. This distinction arises from the potential for deuterium fusion in their massive cores, which can increase temperatures and alter their chemical composition. Therefore, we identify two possible classifications for exoplanets within this mass range: (1) Exoplanets with masses between $1 M_\mathrm{J}$ and $13 M_\mathrm{J}$, described by the isentropic model for their hydrogen and helium profiles, and (2) exoplanets with masses between $13 M_\mathrm{J}$ and $30 M_\mathrm{J}$ which could be either heavy gas giants or brown dwarfs. We expect the model adopted in this work to also apply to heavy gas giants, albeit with potential adjustments to the entropy benchmark values. While our study method does not directly address brown dwarfs, we anticipate similar phenomenology regarding superheavy DM capture and BH formation. However, more detailed profile modeling will be required in future work to refine sensitivity within the DM parameter space. For simplicity, we refer to all objects in the mass range of $1 M_\mathrm{J}$ to $30 M_\mathrm{J}$ as ``exoplanets" throughout this paper.

To model the density profiles of exoplanets with different masses, we assume they follow the same radial density profile as Jupiter, scaled by a consistent factor of the mass ratio.\footnote{We assume that exoplanets lack a core in our analysis and defer addressing complexities arising from varying core sizes among exoplanets of different masses to future work, contingent on the development of more accurate models for cored planetary profiles. We found that simulations conducted with and without a core show no significant differences in the final results under our Jupiter benchmark.} The radii of gas planets do not vary significantly with their masses; therefore, we set the radius of all exoplanets to equal to Jupiter's radius $R_{\rm J}$. We then use the density-temperature relation shown in the left panel of Fig. \ref{F:Density_Temperature_Jupiter} to obtain the radial temperature profile for the chosen entropy values. We assume that all exoplanets are sufficiently old for bulk cooling to have occurred and apply the same isentropic assumption used for Jupiter-sized exoplanets. For an exoplanet with mass of $13 M_\mathrm{J}$, we assume the isentropic contour of $S=7\, k_\mathrm{B} \, \mathrm{e^{-1}}$. For an exoplanet with a mass of $30 M_\mathrm{J}$, we consider two cases with isentropy values of $7\, k_\mathrm{B} \, \mathrm{e^{-1}}$ and $10 \, k_\mathrm{B} \, \mathrm{e^{-1}}$, and perform all analyses for these two isentropic contours. For the chemical composition, we assume all exoplanets in our analysis consist solely of hydrogen and helium, with a mass fraction parameter $Y=0.275$ similar to Jupiter.

\subsection{Methods for Observing Exoplanets}
In this section, we discuss the observational methods for exoplanets in current and upcoming observations. The detection methods depend on the location of the exoplanets, typically classified into two categories: exoplanets within exo-solar systems and rogue exoplanets. Exoplanets bound to host stars can be detected by observing changes in the stars’ motion caused by the gravitational pull of the orbiting planet. Host stars can be tracked using Doppler spectroscopy, which measures their radial velocities through shifts in their absorption spectra~\cite{1952Obs....72..199S}. The radial velocity method has been used to discover more than $1000$ exoplanets. Another method, astrometric measurement, detects the motion of host stars on the plane of the sky with high precision, as demonstrated by the Gaia mission. Gaia’s astrometric method is well-suited for detecting giant planets, including the gas giants considered in this study~\cite{2010exop.book.....S}. The Gaia DR3 catalog has detected $214$ exoplanet candidates, and about $7\times10^4$ exoplanets with mass between $1-15 M_{\rm J}$ are expected to be detected by Gaia astrometry over a $10$-${\rm years}$ observation period~\cite{2023A&A...674A...1G, 2023A&A...674A..10H,2014ApJ...797...14P}.

Exoplanet motion can also be detected via the transit photometry method, which tracks changes in the brightness of the host star. When an exoplanet passes in front of its host star, it partially obscures the starlight, creating a characteristic dip in the observed brightness. By analyzing these dips, researchers can identify exoplanets, estimate their sizes, and determine their orbital parameters. This method, known as the transit method, has been implemented in searches such as the Transiting Exoplanet Survey Satellite (TESS)~\cite{2015JATIS...1a4003R}, which has detected more than $7000$ candidate objects of interest to date. Gaia has also detected $41$ transiting exoplanet candidates, two of which have been further confirmed through radial velocity measurements~\cite{2022A&A...663A.101P}. The upcoming Roman mission is expected to detect $\mathcal{O}(10^5)$ transiting planets, with most of them being giant planets~\cite{2023ApJS..269....5W}. For exoplanets located far from host stars, the transit photometry method is less applicable due to the decreased probability of a transit being observed. However, the direct imaging method can be used to detect their existence and measure their surface temperatures. Direct imaging uses a coronagraph to block the light of the exoplanet's host star and directly detect the infrared signature of orbiting exoplanets. A handful of exoplanets have been discovered using direct imaging thus far, mostly around young stars, but the Roman Space Telescope will include a technology demonstration of the first space-based coronagraph instrument, which will pave the way for further development of space-based direct imaging with future telescopes, such as the Habitable Worlds Observatory.

Exoplanets can also be detected through gravitational lensing measurements, a method that is also useful for identifying rogue planets that do not have a host star~\cite{1992ApJ...396..104G}. In the planetary mass range, this phenomenon is known as microlensing, which occurs when a planet’s mass bends the light from a background source, producing a temporary magnification effect in the observed light curve (see~\cite{2012ARA&A..50..411G} for a recent review). The OGLE project has detected about $100$ exoplanets using microlensing method~\cite{2015AcA....65....1U}, and the Gaia mission has detected $363$ candidate events~\cite{2023A&A...674A..23W}. Regarding future microlensing observations, detecting rogue planets via microlensing is one of the primary goals of the Roman telescope, and $\mathcal{O}(10^3)$ rogue planets are expected to be detected towards the direction of the galactic bulge~\cite{2023AJ....166..108S}. The detection of exoplanets near the Galactic Center region is expected to offer significant potential for studying DM interactions.


\section{Dark Matter Capture in Exoplanets}
\label{S:CaptureRate}
In this section, we review the DM capture process in exoplanets via energy loss through scattering with the planet medium. As DM particles traverse an exoplanet, they scatter off its constituents, and if sufficient kinetic energy is lost, they become gravitationally bound. The time evolution of the number of DM particles, $N_\chi$, within exoplanets for non-annihilating DM is
\begin{equation}
\frac{dN_{\chi}}{dt} = C
\label{E:DM_Number_Rate} 
\end{equation}
where $C$ is the DM capture rate. Note that DM particles can also be captured through scattering with other DM particles already bound to the exoplanet, a phenomenon known as self-capturing \cite{Zentner:2009is}.
Since the focus of the current study is the complementarity between exoplanet searches and terrestrial direct detection experiments, we assume DM particles are collisionless with vanishing self-capture cross sections. Under these circumstances, the total number of captured DM particles increases proportionally with time $N_\chi(t) = C \, t$.

DM particles can undergo multiple scatterings as they traverse exoplanets, which becomes particularly relevant when the scattering cross-section is large and repeated energy loss is required for gravitational capture. The single scattering case is originally studied in~\cite{Griest:1986yu,1987ApJ...321..560G, Gould:1987ir, Gould:1991hx}, and the multiple scattering case is studied in~\cite{Albuquerque:2000rk, Bramante:2017xlb, Bramante:2019fhi}. We follow~\cite{Acevedo:2020gro} for the details of the capture calculation. For heavy DM with $m_{\chi} \gg 1 {\rm GeV}$, DM particles travel in a straight line without changing direction once they fall into the exoplanets and collide with the substances of exoplanets, primarily hydrogen and helium as is described in Sec. \ref{S:ExoplanetStructure}. In this case, the propagation distance of an incident angle $\alpha$ relative to the normal of the planet's surface is given by $2R\cos{\alpha}$ for a planet with radius $R$. The expected number of DM scatterings $\langle \tau_{j} \rangle$ on element $j$ along the trajectory can be expressed as a function of the incident angle $\alpha$,
\begin{equation}
\langle \tau_{j} \rangle \left( \alpha \right) = \int_0^{2R \cos{\alpha}} n_{j}\left(r\right) \sigma_{\chi {j}} \, \mathrm{d} l.
\label{E:NumberOfScattering} 
\end{equation}
Here $\sigma_{\chi j}$ is the scattering cross section of DM with the element $j$. The index $j$ includes contributions from hydrogen and helium. $n_{j}\left(r\right)$ is the number density of the element as a function of the radius $r$ from the center of the exoplanet, with $r = \sqrt{l^2 + R^2 - 2lR \cos{\alpha}}$. 

Denoting the velocity of DM particles in the galactic reference frame as $u$, their velocity shifts to $w = \sqrt{u^2 + v_\mathrm{e}^2}$ as they reach the surface of exoplanets, where $v_\mathrm{e}$ is the escape velocity of the planet. Assuming the initial velocity of DM in the planet is $w_\mathrm{0}$, its kinetic energy becomes $m_{\chi} w_\mathrm{\tau}^2/2 = \prod_{{j}} \prod_{i=1}^{\langle \tau_{j} \rangle} \left(1-\beta_{j} \cos{\phi_{\mathrm{i}}}^2\right)\left(m_\mathrm{\chi} w_\mathrm{0}^2/2\right)$, after the expected $\langle \tau_{j} \rangle$ times of scatterings. Here, $\beta_\mathrm{j} = 4 m_\mathrm{j} m_\mathrm{\chi}/\left(m_\mathrm{j} + m_\mathrm{\chi}\right)^2$, and $\phi_{\mathrm{i}}$ is the scattering angle after the $i$th scattering. We take $\langle \cos{\phi_{\mathrm{i}}}^2 \rangle = 1$ for the tiny scattering angles, which is valid for $m_\mathrm{\chi} \gg m_{j}$ parameter region of our interest.\footnote{We note that a different assumption, $\langle \cos{\phi_{\mathrm{i}}}^2 \rangle = 1/2$, is used in some previous literature for the average value of the scattering angle. Given that the DM mass is much heavier than the target masses, we approximate the scattering angle as zero.} Thus the final DM velocity as a function of $\alpha$ is 
\begin{equation}
w_{\tau}\left(\alpha \right) = \left(1-\beta_\mathrm{H}\right)^{\langle \tau_\mathrm{H} \rangle/2}\left(1-\beta_\mathrm{He}\right)^{\langle \tau_\mathrm{He} \rangle/2} w_\mathrm{0}.
\end{equation}
The final velocity is compared to the exoplanet’s escape velocity to evaluate the probability of the DM particle being captured.

In this work, we study two DM models for the nuclear scattering cross section. The first model represents spin-independent scattering with a coherent effect, where the scattering cross section with a nucleus is enhanced by the atomic mass number $A_j$ as
\begin{equation}
\sigma_\mathrm{\chi j}=A_{j}^2 \left(\frac{\mu_{j}}{\mu_\mathrm{p}}\right)^2 \sigma_\mathrm{\chi p}.
\label{E:ScatteringCrossSetionScaled} 
\end{equation}
The $\mu_{j}$ and $\mu_\mathrm{p}$ are the reduced masses of element ${j}$ and nucleon with $m_\mathrm{\chi}$, respectively, and $\sigma_\mathrm{\chi p}$ is the scattering cross section of the DM particle with the nucleon. The second model assumes that the scattering cross section from the nucleus is independent of the mass number as
\begin{equation}
\sigma_{\chi j}= \sigma_\mathrm{\chi p}.
\label{E:ScatteringCrossSetionNotScaled} 
\end{equation}

To calculate the capture rate, we implement the semi-analytic approach outlined in~\cite{Albuquerque:2000rk, Acevedo:2020gro}. This method involves multiplying the incoming DM flux with the capture probability, which serves as an approximation to the full formula for multiple-scattering capture provided in~\cite{Bramante:2017xlb}. We modify the formula to include the gravitational focusing effect, accounting for the influence of the exoplanet’s gravitational potential on the trajectories of incident DM particles~\cite{Garani:2017jcj, Goldman:1989nd, Kim:2021yyo}, which becomes more significant for exoplanets closer to the GC. Additionally, we included a factor of $1/4$ in the rate to exclude the outgoing particle from the capturing process, ensuring consistency with the original equation in~\cite{Bramante:2017xlb, Zentner:2009is, Spergel:1984re, Gould:1987ir},
\begin{equation}
    C = \pi R^2 \frac{\rho_\mathrm{\chi}}{m_\mathrm{\chi}} \langle v_\mathrm{\chi} \rangle \left(1 + \frac{v_\mathrm{e}^2}{2 \sigma_{\rm v}^2}\right) \int_0^{\pi/2} \mathrm{d}\left(\cos{\alpha}^2\right)  \, \int_0^{v_\mathrm{max}} f\left(w, \theta\right) \, \mathrm{d}w \, \mathrm{d}\left(\cos{\theta}\right),
\label{E:CaptureRate}
\end{equation}
where $\rho_\mathrm{\chi}$ and $\sigma_{\rm v}$ are the DM energy density and 1D velocity dispersion, respectively, at the location of the exoplanet; $v_\mathrm{e}$ is the escape velocity from the surface of the exoplanet, and $v_\mathrm{max}\left(\alpha \right)=v_\mathrm{e} \left(1-\beta_\mathrm{H}\right)^{-\langle \tau_\mathrm{H} \rangle/2}\left(1-\beta_\mathrm{He}\right)^{-\langle \tau_\mathrm{He} \rangle/2}$. The integral term represents the capture probability, and the flux-normalized velocity distribution function is defined as
\begin{align}
f\left(w, \theta\right)=\frac{\left(w^2-v_\mathrm{e}^2\right)^{\frac{3}{2}}}{\bar{N}} \, e^{-\frac{w^2-v_\mathrm{e}^2+v_\mathrm{r}^2+2 v_\mathrm{r} \sqrt{w^2-v_\mathrm{e}^2}\cos{\theta}}{2 \sigma_{\rm v}^2}}
\Theta \left(v_{\rm e h}-\sqrt{w^2-v_\mathrm{e}^2+v_\mathrm{r}^2+2 v_\mathrm{r} \sqrt{w^2-v_\mathrm{e}^2}\cos{\theta}}\right) 
\Theta\left(w-v_\mathrm{e} \right).
\label{E:FluxNormalizedDestributionFunction}
\end{align}
In this equation, $v_\mathrm{r}$ is the velocity of the exoplanet in the galactic rest frame, $\theta$ is the angle between the exoplanet and the DM particle in the galactic rest frame, $v_{\rm e h}$ is the galactic escape velocity at the location of the exoplanet, and $\bar{N}$ is the normalization factor determined by ensuring that the integral $\int_0^{\infty} f\left(w, \theta\right)\, \mathrm{d}w \, \mathrm{d}\left(\cos{\theta}\right)$ equal to unity. 

In Eq. \ref{E:CaptureRate}, the velocity integral can be simplified by removing the step functions as follows:
\begin{equation}
\int \mathrm{d}w \, \mathrm{d}\left(\cos{\theta}\right) =
\left\{ \begin{array}{ll} 
\int\limits_{v_\mathrm{e}}^{v_{\rm max}} \mathrm{d}w \int\limits_{-1}^{1} \mathrm{d}\left(\cos{\theta}\right) & 
v_\mathrm{e} \leq v_{\rm max} < w_* \\
\int\limits_{v_\mathrm{e}}^{w_*} \mathrm{d}w \int\limits_{-1}^{1} \mathrm{d}\left(\cos{\theta}\right) + \int\limits_{w_*}^{v_{\rm max}} \mathrm{d}w \int\limits_{-1}^{c_*} \mathrm{d}\left(\cos{\theta}\right) & 
w_* \leq v_{\rm max} < w'_* \\
\int\limits_{v_\mathrm{e}}^{w_*} \mathrm{d}w \int\limits_{-1}^{1} \mathrm{d}\left(\cos{\theta}\right) + \int\limits_{w_*}^{w'_*} \mathrm{d}w \int\limits_{-1}^{c_*} \mathrm{d}\left(\cos{\theta}\right) & 
w'_* \leq  v_{\rm max}
\end{array} \right. .
\end{equation}
Here, $c_*$, $w_*$, and $w'_*$ are defined as follows:
\begin{align}
c_* = \frac{v_{\rm eh}^2 - \left(w^2 - v_\mathrm{e}^2\right) - v_\mathrm{r}^2}{2 v_\mathrm{r} \sqrt{w^2 - v_\mathrm{e}^2}}, ~~
 w_* = \sqrt{\left(v_\mathrm{eh} - v_\mathrm{r}\right)^2 + v_\mathrm{e}^2}, ~~
w'_* = \sqrt{\left(v_\mathrm{eh} + v_\mathrm{r}\right)^2 + v_\mathrm{e}^2}.
\end{align}
Notably, $v_{\rm max}$ is a function of $\alpha$, and one can integrate $\alpha$ in the last step. $\langle v_\mathrm{\chi} \rangle$ is the average velocity of DM particles relative to the exoplanet at the given galactic distance, given by
\begin{equation}
    \langle v_\mathrm{\chi} \rangle = \frac{ \int \limits_{0}^{\infty}  \mathrm{d} u \, \mathrm{d} (\cos{\theta}) \, 2\pi u^3 \mathrm{e}^{-\frac{u^2 + v_\mathrm{r}^2+2 u v_\mathrm{r} \cos{\theta}}{2\sigma_{\rm v}^2}} \Theta(v_\mathrm{eh} - \sqrt{u^2 + v_\mathrm{r}^2+2 u v_\mathrm{r} \cos{\theta}})}{\left(2\sigma_{\rm v}^2 \pi\right)^{3/2}\left[\mathrm{erf}(\frac{v_\mathrm{eh}}{\sqrt{2}\sigma_{\rm v}})-\frac{2}{\sqrt{\pi}}\frac{v_\mathrm{eh}}{\sqrt{2}\sigma_{\rm v}}\mathrm{e}^{-\frac{v_\mathrm{eh}^2}{2\sigma_{\rm v}^2}}\right]},
\end{equation}
where the denominator is the normalization factor of the truncated Maxwell-Boltzmann velocity distribution in the exoplanet's reference frame.

We study the capture rate for three benchmark exoplanets with masses of $1 M_\mathrm{J}$, $13 M_\mathrm{J}$, and $30 M_\mathrm{J}$. The exoplanets are assumed to be located at two different distances from the Galactic Center, with $d = 1 \, \mathrm{kpc}$ and $d = 8 \, \mathrm{kpc}$. As discussed in Sec. \ref{S:ExoplanetStructure}, we assume that all exoplanets, aside from their overall mass, follow the same density profile (without a core) as that of Jupiter in our Solar System. Similarly, we assume that all exoplanets have the same radius as Jupiter, $R = R_{\rm J}$, and that their chemical composition consists of hydrogen and helium with a mass fraction of $Y = 0.275$. The escape velocities at the surfaces of the exoplanets with masses $1 M_\mathrm{J}$, $13 M_\mathrm{J}$, and $30 M_\mathrm{J}$ are $v_\mathrm{e} = 60.4\,\mathrm{km \, s^{-1}}$, $v_\mathrm{e} = 217.7 \,\mathrm{km\,s^{-1}}$, and $v_\mathrm{e} = 330.7\,\mathrm{km\,s^{-1}}$, respectively. We assume the exoplanet's velocity in the galactic frame is determined by the rotation curve of the Milky Way galaxy, modeled with an NFW profile~\cite{Navarro:1996gj}. 

The details of the Milky Way DM halo, the baryonic matter distribution, as well as the derived rotation curve, velocity dispersion, and galactic escape velocity are provided in Appendix \ref{A:Velocities}. At a distance of $d = 8\,\mathrm{kpc}$ on the galactic plane, the velocity of the exoplanet is $v_\mathrm{r} = 240.6\,\mathrm{km\,s^{-1}}$, the galactic escape velocity is $v_\mathrm{eh} = 588.6\,\mathrm{km\,s^{-1}}$, the DM 1D velocity dispersion is $\sigma_\mathrm{v} = 175.1\,\mathrm{km\,s^{-1}}$, and the average DM velocity is $\langle v_\mathrm{\chi} \rangle = 356.9\,\mathrm{km\,s^{-1}}$. Similarly, at $d = 1\,\mathrm{kpc}$, we find $v_\mathrm{r} = 157.1\,\mathrm{km\,s^{-1}}$, $v_\mathrm{eh} = 725.2\,\mathrm{km\,s^{-1}}$, $\sigma_\mathrm{v} = 173.2 \,\mathrm{km \,s^{-1}}$, and $\langle v_\mathrm{\chi} \rangle = 312.5 \,\mathrm{km \, s^{-1}}$. For the DM density, we find $\rho_\mathrm{\chi} = 0.38\,\mathrm{GeV\,cm^{-3}}$ at the local distance $d = 8\,\mathrm{kpc}$, and $\rho_\mathrm{\chi} = 5.35\,\mathrm{GeV\,cm^{-3}}$ at $d = 1\,\mathrm{kpc}$.

\begin{figure}[t]
    \centering
    \includegraphics[width=0.5\columnwidth]{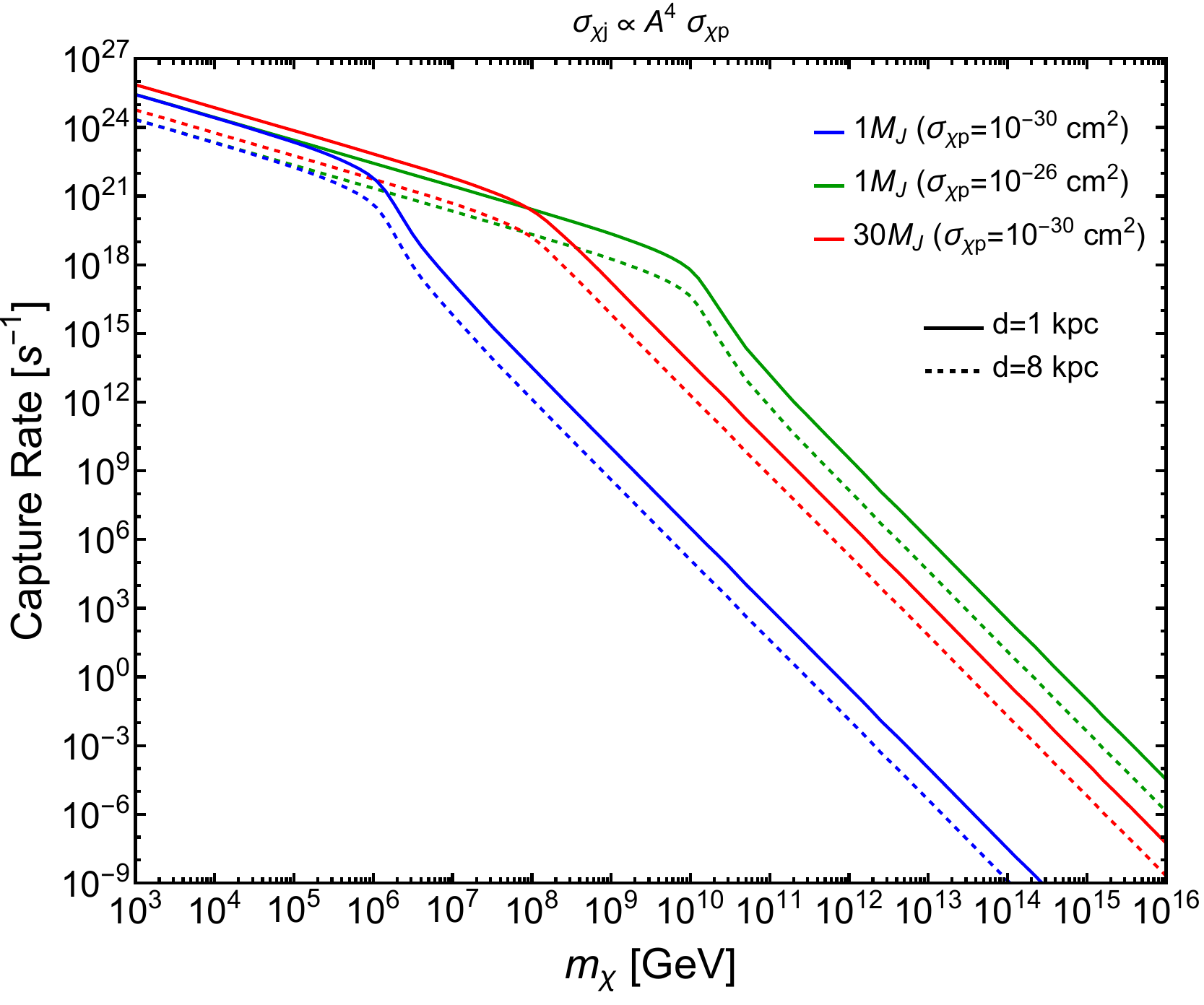}
    \caption{Capture rate versus DM mass for the spin-independent coupling model with exoplanets located at different distances from the galactic center $d=1 \, \mathrm{kpc}$ (solid) and $d=8 \, \mathrm{kpc}$ (dashed). The blue curves represent the capture rate for an exoplanet mass of $1M_\mathrm{J}$ and a cross section of $\sigma_\mathrm{\chi p}=10^{-30} \, \mathrm{cm^2}$, the green curves are for a mass of $1 M_\mathrm{J}$ and a cross section of $\sigma_\mathrm{\chi p}=10^{-26} \, \mathrm{cm^2}$, and the red curves are for a mass of $30 M_\mathrm{J}$ and a cross section of $\sigma_\mathrm{\chi p}=10^{-30} \, \mathrm{cm^2}$.}
    \label{F:CaptureRate}
\end{figure}

Fig.~\ref{F:CaptureRate} shows the capture rate for exoplanets with varying masses and galactic locations, based on the model incorporating the coherent effect in Eq.~\eqref{E:ScatteringCrossSetionScaled}. The blue curves represent the capture rate for an exoplanet mass of $1 M_\mathrm{J}$ with the cross section of $\sigma_\mathrm{\chi p} = 10^{-30} \, \mathrm{cm}^2$, the green curves represent an exoplanet mass of $1M_\mathrm{J}$ and $\sigma_\mathrm{\chi p} = 10^{-26} \, \mathrm{cm}^2$, and the red curves for an exoplanet mass of $30 M_\mathrm{J}$ and $\sigma_\mathrm{\chi p}=10^{-30} \, \mathrm{cm}^2$. For all colors, the solid curves indicate that the exoplanet is located at a distance of $d = 1\, \mathrm{kpc}$, and the dashed curves indicate a location of $d = 8\, \mathrm{kpc}$. The capture rate decreases with increasing DM mass for two reasons: firstly, the flux of DM particles incident on exoplanets is inversely proportional to $m_\chi$; secondly, heavier DM particles do not lose sufficient kinetic energy to become gravitationally bound to the exoplanet after a fixed number of scatterings (for the chosen cross section value). Therefore, nearly all DM particles are captured for the benchmarks in the lower DM mass regions in Fig.~\ref{F:CaptureRate}, whereas the capture rate decreases more rapidly in the higher mass regions due to the additional kinematic suppression. For a fixed exoplanet mass, the capture rate increases with $\sigma_{\chi\textrm{p}}$ due to the higher number of scatterings $\langle \tau_{j} \rangle$ in Eq.~\eqref{E:NumberOfScattering}. Similarly, $\langle \tau_{j} \rangle$ and the resultant capture rate increase with the exoplanet’s mass because of the higher target number density. The larger exoplanet mass also enhances the capture rate through a stronger gravitational focusing effect. 

We remark that the evaporation of the captured DM abundance does not occur within the parameter space of interest. One can estimate the highest DM mass that can escape from the center of exoplanets by thermalizing with the central temperature and obtaining a kinetic energy $\frac{3}{2} \, T(0) = \frac{1}{2} m_\mathrm{\chi} v_\mathrm{e}^2$ \cite{Gould:1989tu}. For exoplanets with escape velocities of $60-500 \, {\rm km \, s^{-1}}$ and core temperatures of $10^4-10^5 \, \mathrm{K}$, the evaporation mass is less than $1 \, {\rm GeV}$. Given that the BH formation process occurs at large enough DM masses, the evaporation loss is negligible in our simulations. Furthermore, we calculated the capture rate for a Jupiter-sized exoplanet with a core, adopted from the same reference \cite{MARLEY2014743}, and found no significant change in the capture rate.


\section{Black Hole Formation in Exoplanets}
\label{S:BHFormationCondition}
In this section, we analyze the formation of BHs in the center of exoplanets after sufficient DM abundance is captured. The crucial condition for BH formation relies on the time scales associated with the evolution of the DM abundance within the exoplanet’s interior. We derive the time scales for capture, thermalization, drifting, and gravitational collapse. We then compare the total time required with the age of exoplanets to outline the relevant parameter space. 

After DM particles become gravitationally bound to the exoplanet through the capture process discussed in the previous section, they continue to thermalize with SM particles and drift toward the exoplanet’s central regions. Eventually, the DM clump at the center may collapse into a BH. We model the BH formation time, $t_{\rm form}$, as the sum of all relevant processes from the initial capture to the final collapse
\begin{equation}
t_{\rm form} = t_{\rm cap} + t_{\rm th}^{\rm I} + t_{\rm th}^{\rm II} + t_{\rm drift} + t_{\rm col}.
\label{E:totalformationtime}
\end{equation}
We note that this represents a conservative assumption for the BH formation process, as the time scales in Eq.~\eqref{E:totalformationtime} may overlap when multiple processes occur simultaneously.
The $t_{\rm cap}$ is the time required to capture a sufficient DM abundance. Given the capture rate from the last section, the capture time is expressed as $t_{\rm cap}=M_{\rm crit}/(m_\mathrm{\chi} C)$, where $M_{\rm crit}$ is the critical mass required for BH formation, which will be introduced later in this section. The other time scales, $t_{\rm th}^{\rm I}$, $t_{\rm th}^{\rm II}$, $t_{\rm drift}$, and $t_{\rm col}$, correspond to the first and second thermalization times, the drift time, and the collapse time, respectively, which, following \cite{Acevedo:2020gro}, will be discussed in detail below.


\subsection{Thermalization Time}
After being captured by exoplanets, DM particles initially follow trajectories larger than the size of the exoplanets. The DM orbit shrinks as it passes through the exoplanet, losing kinetic energy through scattering. Therefore the thermalization of DM depends on the DM mass and the scattering cross section with nucleons. We classify the thermalization process into two stages. The first thermalization refers to the time required for a captured DM particle’s orbit to shrink to the size of the exoplanet, confining the DM within it. The second thermalization represents the subsequent stage, during which the DM particle becomes fully enclosed within the exoplanet and gradually sinks toward its center~\cite{Kouvaris:2010jy}.

The time required for the first thermalization is determined based on the total energy loss needed and the rate of energy loss through scattering. The energy loss rate can be determined with the average energy loss accumulated during each crossing $\langle \Delta E\rangle_{\rm}$ and the average orbital period $\Delta t$~\cite{Acevedo:2020gro, Acevedo:2019gre, Kouvaris:2010jy}, 
\begin{equation}
    \frac{dE_{\rm}}{dt}=-\frac{\langle \Delta E\rangle_{\rm}}{\Delta t}.
\label{E:FirstThermalizaton}
\end{equation}
The average energy loss of a DM particle during each complete passage through the exoplanet is given by
\begin{equation}
    \langle \Delta E\rangle = \frac{\langle \tau \rangle}{R}\int_0^R \Delta E \, \mathrm{d}r, 
\end{equation}
where $R$ is the exoplanet's radius, $\langle \tau \rangle$ is the total number of scatterings with hydrogen and helium in Eq.~\eqref{E:NumberOfScattering}, $\Delta E = 2 \bar{\mu} K / m_\mathrm{\chi}$, $\bar{\mu}$ is the average reduced mass, and $K$ is the initial kinetic energy of the DM particle. As discussed in Sec. \ref{S:ExoplanetStructure}, the average mass of elements is $\bar{m} = 1.18\, \mathrm{GeV}$ with $Y = 0.275$, and the average reduced mass defined as $\bar{\mu} \equiv \bar{m} m_\mathrm{\chi} / (\bar{m} + m_\mathrm{\chi})$ is calculated accordingly. Taking the model with the coherent effect as an example, for a Jupiter-sized exoplanet, we find $\langle \tau \rangle \simeq 7\times 10^8$ accumulated for hydrogen and helium according to Eq. \eqref{E:NumberOfScattering}, with $\alpha = 0$, $\sigma_\mathrm{\chi p} = 10^{-27} \, \mathrm{cm}^2$, and $m_\mathrm{\chi} = 10^7 \, \mathrm{GeV}$. We express the kinetic energy in terms of the total energy and potential energy as $K = E -  \Phi(r)$ for convenience in later calculations, where the rest mass $m_\chi$ is excluded from $E$. The potential energy at a distance $r$ from the center of the exoplanet is given by $\Phi\left( r\right) =-\int_r^{\infty} G m_\mathrm{\chi} M\left(r'\right)/r'^2 \, \mathrm{d}r'$, where $M(r)$ is the enclosed mass within radius $r$, and $G$ is the gravitational constant. The average orbital period of DM particles can be approximated with 
\begin{equation}
    \Delta t = 2 \pi \sqrt{\frac{a^3}{G M}}, 
\end{equation}
where $a$ is the semi-major axis of the orbit and is related to the total energy $E$ by $a=-G M m_\mathrm{\chi}/2 E$. After replacing $\langle \Delta E \rangle$ and $\Delta t$ in Eq.~\ref{E:FirstThermalizaton}, we integrate the energy from its initial value to its final value. The initial energy corresponds to the time when the DM particle is first bound to the exoplanet. The velocity of a DM particle after capture is initially less than the escape velocity, $E \leq 0$. To avoid singularity, we ignore the case of zero initial energy and roughly assume that the initial total energy corresponds to the energy after one scattering, $E^{\rm i} \simeq -\bar{\mu}^2 v_\mathrm{e}^2/\bar{m} = -2\bar{\mu}^2 G M / \bar{m} R$. We roughly assume that the final energy is equivalent to the major axis of the elliptical orbit becoming the size of the planet, at which point the total energy will be $E^{\rm f}=-GM m_\mathrm{\chi}/2 R$. Finally, the first thermalization time will be:
\begin{equation}
   t^{I}_{\rm th}= \int^{E^{\rm f}}_{E^{\rm i}} \frac{- \mathrm{d}E \, \Delta t}{\langle \Delta E\rangle }.
\end{equation}
For example, the first thermalization in a Jupiter-sized exoplanet takes about $3 \, {\rm years}$ to complete, assuming $m_\mathrm{\chi}=10^9 \, \mathrm{GeV}$ and $\sigma_\mathrm{\chi p}=10^{-27} \, \mathrm{cm}^2$ with the coherent scattering effect.

After DM particles become confined within exoplanets, their velocity remains significantly larger than the thermal velocity of the exoplanet’s constituents. The second thermalization occurs through additional scatterings that result in energy loss, eventually reducing the DM velocity $v_\chi$ to below the constituents' velocity $v$, given that $m_\mathrm{\chi} \gg \bar{m}$.\footnote{Note the difference in notation used in this section for velocities compared to that in Sec. \ref{S:CaptureRate}.} We assume the thermal velocity of the elements in exoplanets is $v=\sqrt{3 T/\bar{m}}$, which varies at different radii. The scenario where $v_\mathrm{\chi} \gg v$ is referred to as the ``inertial" regime, while the scenario where $v_\mathrm{\chi} \ll v$ is referred to as the ``viscous" regime. The second thermalization starts in the inertial regime and transitions to the viscous regime. To calculate the second thermalization time, we identify the energy loss rate
\begin{equation}
   \frac{dE}{dt}=- \Gamma \, \langle \delta E \rangle,
\label{E:EnergyLoss}
\end{equation}
where the scattering rate is $\Gamma = \rho\left(r\right) \bar{\sigma}_{\chi j} v_{\rm rel}/\bar{m}$ with $\rho$ being the density of the exoplanet, and the average energy loss per scattering is $\langle \delta E \rangle = \bar{\mu}^2 v_\mathrm{\chi}^2/\bar{m}$. We can use the average atomic mass, $\bar{A}=1.26$ in our model, and thus the average scattering cross section scales as $\bar{\sigma}_{\chi j}=\bar{A}^2 \bar{\mu}^2 \sigma_\mathrm{\chi p} / \mu_\mathrm{p}^2$ in the model with the coherent effect in Eq.~\eqref{E:ScatteringCrossSetionScaled}. The relative velocity $v_{\rm rel}$ in the rate $\Gamma$ is determined by $v_\chi$ in the inertial regime and by $v$ in the viscous regime. Due to the lower energy loss rate in the viscous regime, the viscous regime dominates the time scale of the second thermalization stage. Therefore, we conservatively assume the relative velocity to be $v_{\rm rel} \simeq v = \sqrt{3T/\bar{m}}$ throughout the entire second thermalization process. Thus, the second thermalization time can be approximated by considering only the viscous regime. By substituting $v_\mathrm{\chi}^2=2\left(E-\Phi \left(r \right) \right)/m_\mathrm{\chi}$, we can solve the DM particle energy as a function of time 
\begin{equation}
   E=\Phi\left(r \right)+\left(E^{\rm i}-\Phi\left(r \right)\right)\mathrm{e}^{- \frac{2 \bar{\mu}^2}{m_\mathrm{\chi} \bar{m}}\Gamma \, t}.
\end{equation}
The initial energy $E^{\rm i}$ can be approximated as the DM energy at the onset of the viscous regime, where $v_\mathrm{\chi} \simeq v = \sqrt{3 T/\bar{m}}$, resulting in $E^{\rm i}=\Phi\left(r \right)+\left( 3 m_\mathrm{\chi}/2 \bar{m}\right)T$. The final energy $E^{\rm f}$ is given by the thermalization condition, $v_{\chi}= \sqrt{3T/m_\mathrm{\chi}} \ll v$, yielding $E^{\rm f}=\Phi\left(r \right)+3T/2$. 
Note that the second thermalization can occur at any location within the exoplanet, which means the thermalization time scale varies at different radii. In our analysis, we use the average values for the temperature and density from the exoplanet profile.
The second thermalization time is expressed as
\begin{equation}
   t_{\rm th}^{\rm II}= \frac{\bar{m}^2 m_\mathrm{\chi}}{ 2 \bar{\mu}^2 \bar{\sigma}_\mathrm{\chi j}}  \left(\frac{\int_0^R \frac{\mathrm{d}r}{\rho\left(r\right)\,\sqrt{3 T\left(r\right)/\bar{m}}}}{R}\right) \ln{\frac{m_\mathrm{\chi}}{\bar{m}}}.
\end{equation}
For a Jupiter-sized planet with $m_\mathrm{\chi}=10^9 \, \mathrm{GeV}$ and $\sigma_\mathrm{\chi p}=10^{-27} \, \mathrm{cm}^2$, the second thermalization completes in about $5 \, {\rm years}$ in the spin-independent scattering model.


\subsection{Drift Time}
Following the thermalization stage, DM particles migrate toward the center of exoplanets, forming a DM clump. If the scattering cross section is excessively large, the drift time determined by the viscous drag force could exceed the age of the exoplanet. Consequently, for BH formation to proceed, the drift time must be sufficiently short to allow efficient DM accumulation at the center \cite{Bramante:2019fhi, Acevedo:2020gro, Gould:1989gw}. We conservatively assume that the drifting stage happens after the completion of the second thermalization, and the total drifting distance spans from the exoplanet’s surface to its center, although the actual drift timescale may be shorter than our approximation. The equilibrium equation below shows the balance between gravity and the viscous drag force acting on DM particles, with contributions from each chemical component $j$ of the exoplanet, at a radius $r$ from the center of the exoplanet
\begin{equation}
   \frac{G M\left(r\right) m_\mathrm{\chi}}{r^2} = v_{\rm drift} \left[\sum \limits_{j} n_{j}\left(r\right) \, m_{j} \, \langle \sigma _{\chi j} v_{\rm th}\rangle \right].
\end{equation}
The drifting velocity is $v_{\rm drift}={\rm d}r/{\rm d}t$, and thermal velocity of $j$ is $v_{\rm th}=v_{j}=\sqrt{3T/m_{j}}$. By solving this equation, we obtain the drift time as 
\begin{equation}
   t_{\rm drift}=\frac{1}{G m_{\chi}}\left[\sum \limits_{j}  \sigma _{\chi j} \int_0^{R}\frac{n_{j}\left(r\right) \, \sqrt{3 m_{j} \, T\left( r\right) }}{M\left(r\right)}r^2 \mathrm{d}r \right].
   \label{E:DriftTime}
\end{equation}
As expected, the drift time increases with the scattering cross section, as indicated by the expression above. Taking $m_\mathrm{\chi} = 10^9 \, \mathrm{GeV}$, the drift time is about $10$ minutes for $\sigma_\mathrm{\chi p} = 10^{-27} \, \mathrm{cm}^2$ in Eq.~\eqref{E:ScatteringCrossSetionScaled}, and about $2$ years for $\sigma_\mathrm{\chi p} = 10^{-22} \, \mathrm{cm}^2$ in the spin-independent scattering model.


\subsection{Collapse Time}
The accumulated DM abundance at the center of an exoplanet forms a spherical distribution through virialization, with the radius of the DM sphere determined by the virial theorem, $\langle K\rangle =-\langle \Phi \rangle/2$ \cite{Spergel:1984re, Faulkner:1985rm, Griest:1986yu,Gould:1991hx}. Assuming the energy density of baryonic matter $\rho$ dominates the contributions to the gravitational potential at the center
\begin{equation}
   r_{\rm th}=\sqrt{\frac{9 \, T\left(0\right)}{4 \pi G m_{\chi} \, \rho\left(0\right)} }.
\end{equation}
Here we have assumed that the DM within the sphere is thermally equilibrated.

The DM clump must satisfy several conditions to efficiently collapse into a BH. These include the Jeans instability condition and the Chandrasekhar limit \cite{Acevedo:2020gro}. The Jeans instability condition requires that the sound crossing time must exceed the free-fall time of the DM particle. The speed of sound in the DM sphere is approximately $c_{\rm s}\simeq \sqrt{T\left(0\right)/m_\mathrm{\chi}}$. Consequently, the sound crossing time is given by $t_{\rm s}=r_{\rm th}/c_{\rm s}=3/\sqrt{4\pi G \rho\left(0\right)}$. The DM free-fall time can be calculated by modeling the DM sphere distribution as a collapsing spherical cloud of gas with a homogeneous density of $\rho_\chi^{\rm cl}$ and a radius of $r_{\rm th}$. The free-fall time of a spherically symmetric gas cloud, as a test particle on the surface is pulled toward the center by the gravitational acceleration $4\pi \rho_\chi^{\rm cl} r_{\rm th}^3 G / 3 r^2$, is given by $t_{\rm ff}=\sqrt{3\pi/ (32 G \rho_\chi^{cl})}$. The Jeans instability condition requires $t_{\rm s} > t_{\rm ff}$, which implies that for collapse to occur, the DM density must surpass the central density of the exoplanet, \textit{i.e.}, $\rho_\chi^{\rm cl} > \rho(0)$.

Once the DM clump is formed, virialized, and thermalized at the center of the exoplanet, its density may no longer remain subdominant. As DM continues to accumulate, the total captured DM mass $M_{\rm cap}$ can become comparable to that of hydrogen and helium. Therefore, we refine the virial equation to also accommodate the DM contribution, 
\begin{equation}
   2 \left(\frac{3 T\left(0\right) }{2}\right)=-\left(-\frac{4}{3}\pi r^2 G m_\mathrm{\chi}\rho\left(0\right)-\frac{G M_{\rm cap} m_{\chi}}{r}\right).
\end{equation}
The equation above has a solution if the following condition is satisfied
\begin{equation}
   M_{\rm cap} \geq M_{\rm sg}= \sqrt{\frac{3 T\left(0\right)^3}{\pi G^3 m_\mathrm{\chi}^3 \rho\left(0\right)}}, 
\end{equation}
where $M_{\rm sg}$ stands for the self-gravitational mass. This condition implies that the gravitational potential of the DM sphere surpasses the thermal pressure, and thus the sphere becomes unstable and collapses.

The second DM mass condition corresponds to the Chandrasekhar limit. The accumulated DM mass should be sufficiently large to overcome the quantum degeneracy pressure, which arises from the uncertainty principle for bosonic DM and the Pauli exclusion principle for fermionic DM \cite{McDermott:2011jp, Gresham:2018rqo}. For fermions, the Fermi degeneracy condition is expressed as $M_{\rm cap} \geq M_{\rm ch} \simeq M_{\rm pl}^3 / m_\mathrm{\chi}^2$. For bosons, the condition is $M_{\rm cap} \geq M_{\rm ch} \simeq M_{\rm pl}^2 / m_\mathrm{\chi}$~\cite{McDermott:2011jp}. We neglect modifications to the Chandrasekhar limit arising from DM self-interactions, as this study considers only interactions between DM and nuclei. 

Combining the Jeans instability and the Chandrasekhar limit conditions, the DM clump collapse into a BH when its mass becomes larger than the critical mass
\begin{equation}
   M_{\rm cap} \geq M_{\rm crit}= \mathrm{max}\left\{M_{\rm ch}, M_{\rm sg}\right\}. 
   \label{E:CriticalMass}
\end{equation}
We calculate the time scale gravitational collapse in the following. In contrast to the second thermalization process, a DM particle initially undergoes gravitational infall in the viscous regime as its initial virialized velocity, determined with $K_{\rm i}=G M_{\rm crit } m_\mathrm{\chi}/2 r_{\rm th}$, is smaller than that of the baryonic particles. As the DM particle is accelerated during the infall, it eventually gains sufficient kinetic energy to transition into the inertial regime. Since the DM velocity is much larger in the inertial regime, the collapse during this stage completes rapidly. Therefore, the end of the collapse process is effectively determined by the transition from the viscous to the inertial regime, indicated by $v_\mathrm{\chi}=v=\sqrt{3T\left(0\right)/\bar{m}}$, or equivalently, when the kinetic energy reaches $K_{\rm f}=3T \left(0\right) m_{\chi}/2 \bar{m}$. The changes in kinetic energy, $\Delta K=K_{\rm f} - K_{\rm i}$, of the virialized system equals the total energy loss caused by viscous drag. Therefore, one can use the energy loss rate in the viscous regime in Eq.~\eqref{E:EnergyLoss} to determine the collapse time with $\int_0^{t_{\rm col}} \, dE=\Delta K$. The collapse time is given by
\begin{equation}
   t_{\rm col}=\frac{m_\mathrm{\chi} \bar{m}^2}{2 \bar{\mu}^2 \rho\left(0\right) \bar{\sigma}_{\rm \chi j} \sqrt{3T\left(0\right)/\bar{m}}} \ln{\left[\frac{3T\left(0\right)}{\bar{m}} \frac{r_{\rm th}}{G M_{\rm crit}}\right]}. 
\end{equation}
For instance, for a fermionic DM mass of $m_\mathrm{\chi}=10^9 \, \mathrm{GeV}$ and a scattering cross section of $\sigma_\mathrm{\chi p}=10^{-27} \, \mathrm{cm}^2$, the collapse in the spin-independent scattering model takes about $11 \, {\rm days}$.


\subsection{Time Evolution after Black Hole Formation}
Once the BH forms, its mass evolution is determined by the accretion of baryonic matter and subsequently captured DM, as well as the evaporation due to Hawking radiation \cite{Acevedo:2019gre, Acevedo:2020gro}. We assume the initial BH mass is $M_{\rm BH}^{\rm init}=M_{\rm crit}$, and the mass evolution is 
\begin{equation}
   \frac{dM_{\rm BH}}{dt}= \frac{4 \pi \rho\left(0\right) G^2 M_{\rm BH}^2}{c_{s*}^3}+e_\mathrm{\chi} m_\mathrm{\chi} C-\frac{f}{G^2 M_{\rm BH}^2}. 
\label{E:BlackHoleEvolution}
\end{equation}
The first term corresponds to the Bondi accretion of baryonic matter~\cite{Bondi:1944rnk}. The sound speed at the center of the exoplanet, $c_{s*}$, is approximated as $c_{s*} \simeq \sqrt{T\left(0\right) / \bar{m}}$. The second term represents the accretion of DM abundance captured after the BH formation, with $e_\mathrm{\chi}$ denoting the efficiency factor, which encapsulates the details of DM accretion and is approximated as unity. The last term describes the Hawking radiation~\cite{Hawking:1974rv}, with $f$ being the Page factor. In our simulation of Eq.~\eqref{E:BlackHoleEvolution}, we assume that evaporation occurs exclusively to photons in the geometrical optics limit, for which $f = 1 / (15360 \pi) \, \mathrm{kg \, m^6 \, s^{-5}}$. We will discuss the DM parameter space for BH formation and the associated BH mass evolution in the following section.


\section{Results}
\label{S:Results}

\begin{figure}[t]
    \centering
    \includegraphics[width=0.49\columnwidth]{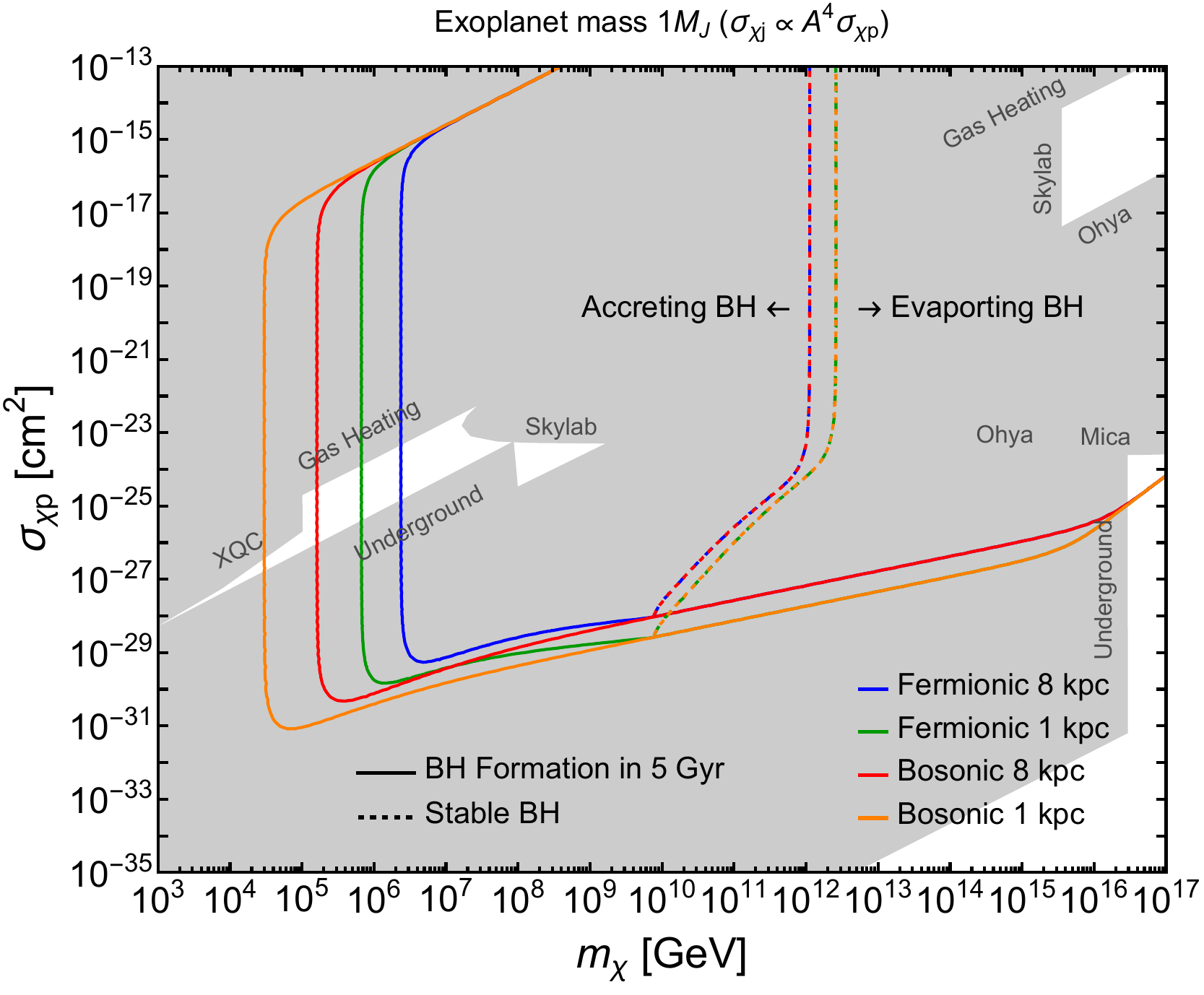} \,\,\,
    \includegraphics[width=0.49\columnwidth]{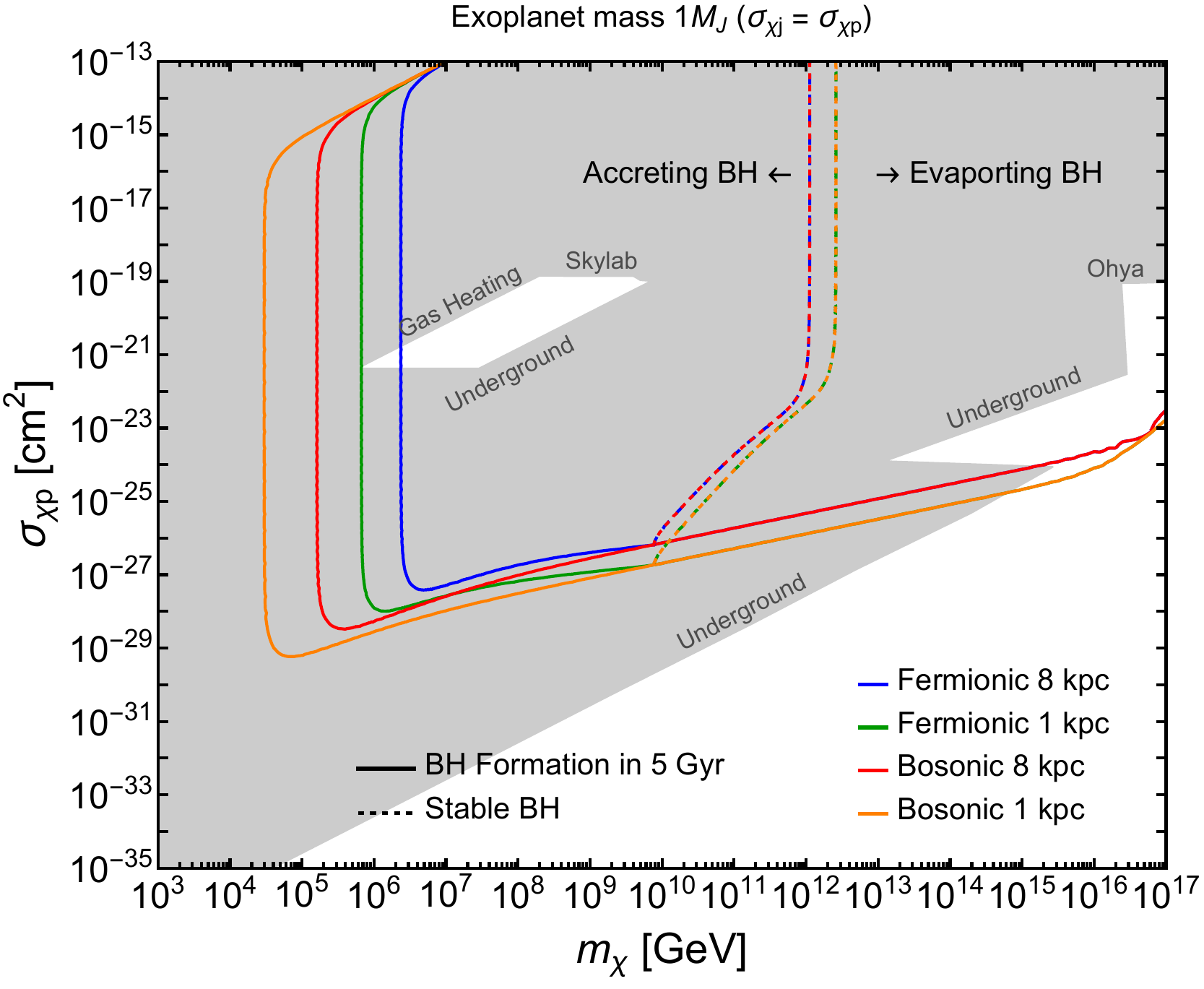}
    \caption{The parameter space where the total time for DM capture and BH formation in an exoplanet of mass $1M_\mathrm{J}$ is less than $5\, \mathrm{Gyr}$, shown for different DM scattering cross section models. Solid curves delineate the boundaries of BH formation regions under various DM models and exoplanet assumptions. Blue and green curves correspond to fermionic DM for exoplanets located at $8\,\mathrm{kpc}$ and $1\,\mathrm{kpc}$ from GC, respectively, while red and orange represent bosonic DM at the same galactic distances. The entropy of the exoplanets is set to $S = 7 \, k_\mathrm{B} \, {\rm e}^{-1}$.  Dashed curves indicate the scenario where BH mass evolution is static. Regions to the left of the dashed curve is dominated by accretion, while regions to the right is dominated by Hawking evaporation. Existing constraints from heavy DM searches are indicated by the gray shaded regions.} 
    \label{F:CrossSectionBoundsScaledNotScaled}
\end{figure}

In this section, we present the viable DM parameter space for DM capture and BH formation on the $\sigma_{\chi {\rm p}}-m_\chi$ plane, for two scattering cross section models. We select benchmark exoplanet candidates characterized by their mass, radius (with $R=R_{\rm J}$ in this work), interior entropy, and location in the Milky Way galaxy. Using these parameters, we calculate the capture rate with Eq.~\eqref{E:CaptureRate} and the corresponding BH formation time with Eq.\eqref{E:totalformationtime}. The formation time is then compared to the exoplanet’s age to identify potential observational signatures.

In Fig.~\ref{F:CrossSectionBoundsScaledNotScaled}, we show the parameter regions where the BH formation time in an exoplanet with a mass of $1M_{\rm J}$ is less than $5~{\rm Gyr}$, representing a typical planetary age. The left panel corresponds to the DM model in Eq.~\eqref{E:ScatteringCrossSetionScaled}, and the right panel represent that in Eq.~\eqref{E:ScatteringCrossSetionNotScaled}. The solid curves denote the boundaries within which the BH formation time satisfies $t_{\rm form} \leq 5 \,\mathrm{Gyr}$. We use different colors to represent DM models and the galactic location of the exoplanet. The blue and green curves represent the fermionic DM case, with the exoplanet located at distances of $8~{\rm kpc}$ and $1~{\rm kpc}$ from GC, respectively. The green curve, representing an exoplanet closer to the GC, probes a larger parameter region compared to the blue curve, which corresponds to a local exoplanet. This enhanced effect is due to the higher capture rate resulting from the increased DM energy density. BH formation for bosonic DM is more viable for smaller DM masses compared to the fermionic DM case, due to the difference in the Chandrasekhar limit in Eq.~\eqref{E:CriticalMass}, which defines the left edge of the contour. The upper and lower edges are less dependent on the spin of the DM particle. In regions with a large scattering cross section above the upper edges, BH formation is prevented due to the stronger viscous drag force, which extends the DM drift time in Eq.~\eqref{E:DriftTime} beyond $5\,\mathrm{Gyr}$. It is worth noting that the drift time depends only on the temperature and density profiles of the exoplanet and not on its distance to the GC. The smaller cross section regions below the lower edges cannot form BH timely because the capture rate is too low for sufficient DM, as required by the critical mass, to accumulate within $5\,\mathrm{Gyr}$. As illustrated in the figure, exoplanets are effective for probing heavy DM mass regions. To show the complementarity of exoplanet probes with existing searches, we include the existing constraints from heavy DM searches, represented by the gray shaded regions, for Mica \cite{Acevedo:2021tbl}, Skylab and Ohya \cite{Bhoonah:2020fys}, gas heating \cite{Bhoonah:2018gjb, Bhoonah:2020dzs}, cosmic microwave background (overlapping with the gas heating bound) \cite{Gluscevic:2017ywp, Bhoonah:2018gjb, Bhoonah:2020dzs}, XQC \cite{Erickcek:2007jv, Bhoonah:2020dzs}, and underground experiment bounds adopted from \cite{Bhoonah:2020dzs, Bhoonah:2018gjb}. Furthermore, we performed the same analysis for a Jupiter-sized exoplanet with a cored density profile and found a change of less than $50\%$ in the bounds.

After the formation of the BH, it may either grow due to accretion and eventually destroy the host exoplanet, or evaporate through Hawking radiation into SM particles. The mass evolution is determined with Eq.~\eqref{E:BlackHoleEvolution}, and we assume the initial mass of the BH is $M_\mathrm{BH}^\mathrm{init}=M_\mathrm{crit}$. According to the relationship between the critical mass and the DM mass, a lighter $m_\chi$ generally leads to the formation of a BH with a larger initial mass, which tends to accrete and become stable. On the other hand, for sufficiently heavy DM masses, the BH evaporates quickly, suggesting that observational signatures arise from the Hawking radiation process. The boundary between the two opposing regimes is defined by the combination of DM mass and cross section that satisfies ${\rm d}M_{\rm BH}/ {\rm d} t=0$ at the formation time. We solve for the DM parameters that result in a static BH mass after formation and plot them as dashed curves for different benchmarks in Fig.~\ref{F:CrossSectionBoundsScaledNotScaled}. BHs formed in the regions to the left of the dashed curve grow via accretion, while those formed to the right of the dashed curve evaporate. The difference between the dashed curves for planets at different distances arises from the contribution of the DM accretion term in Eq.~\eqref{E:BlackHoleEvolution}. In regions with smaller $\sigma_{\chi \rm p}$ values, DM accretion becomes suppressed, and a larger initial BH mass (corresponding to a smaller DM mass) is needed as the curves bend to the left. In the case where the evaporation dominates after BH formation, the evaporation lifetime of the BH is
\begin{equation}
   t_{\rm BH}=\frac{b^{-}}{a_2}\arctanh{\frac{M_{\rm BH}^{\rm init}}{b^-}}-\frac{b^{+}}{a_2}\arctan{\frac{M_{\rm BH}^{\rm init}}{b^+}},
\label{E:BlackHoleTime}
\end{equation}
where $b^{\pm}=\sqrt{\left(a_2\pm a_1\right)/2 l}$, $a_2=\sqrt{\rho\left(0\right)/(960c_*^3)+a_1^2}$, $a_1=e_\mathrm{\chi} m_\mathrm{\chi} C$, and $l=4\pi \rho\left(0\right) G^2/c_*^3$.\footnote{We note that the plus and minus signs of $b^{\pm}$ in our result are opposite to those in \cite{Acevedo:2020gro}.} For example, at $d=1\, \mathrm{kpc}$, for a fermionic DM with a mass of $m_\mathrm{\chi}=10^{16} \, \mathrm{GeV}$ and a scattering cross section of $\sigma_\mathrm{\chi p}=10^{-16} \, \mathrm{cm^2}$ in the left panel of Fig.~\ref{F:CrossSectionBoundsScaledNotScaled}, the evaporation time is $t_{\rm BH} \simeq 10^{-12} \, \mathrm{s}$.

\begin{figure*}[t]
    \centering
    \includegraphics[width=0.49\columnwidth]{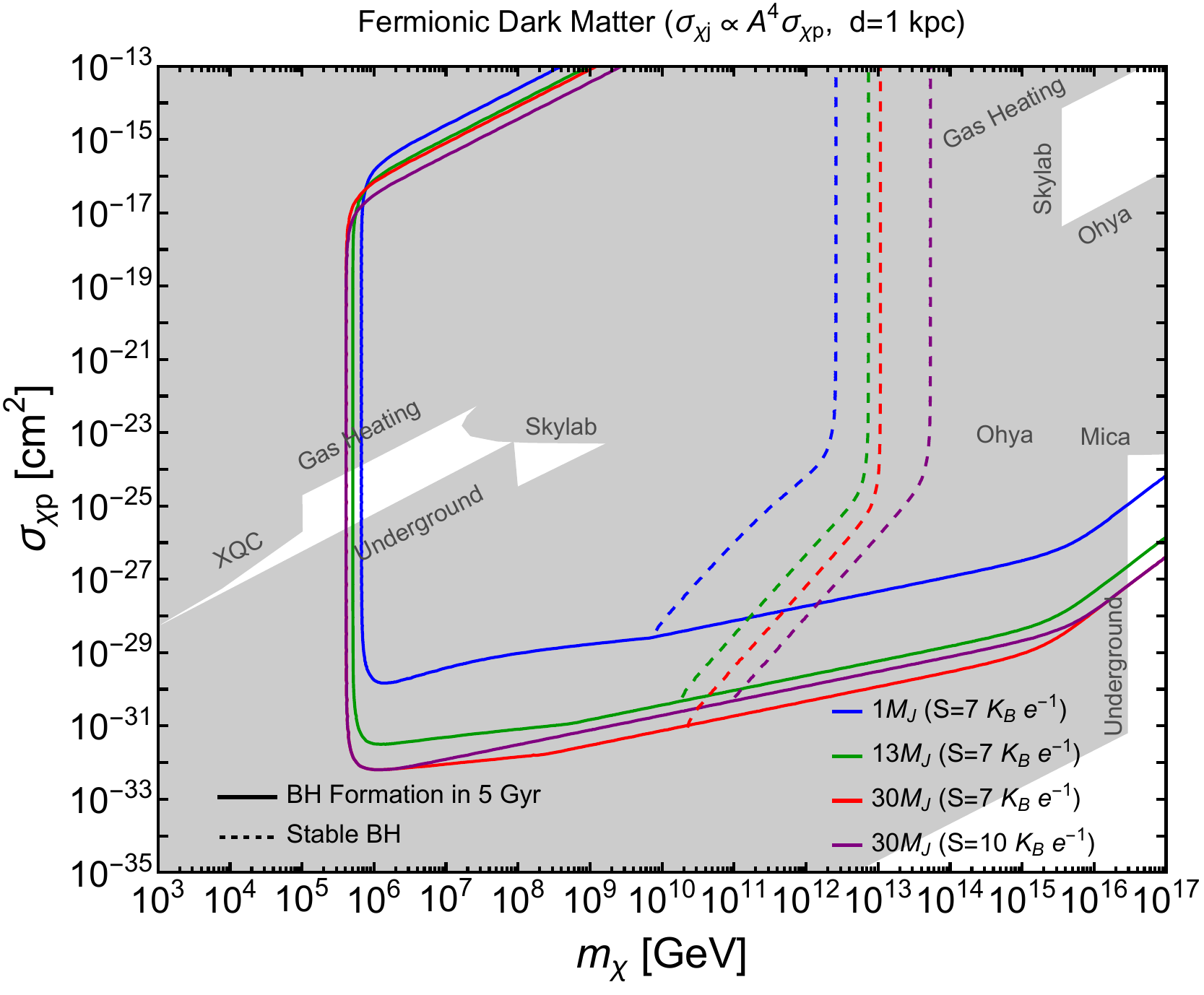}\,\,\,
    \includegraphics[width=0.49\columnwidth]{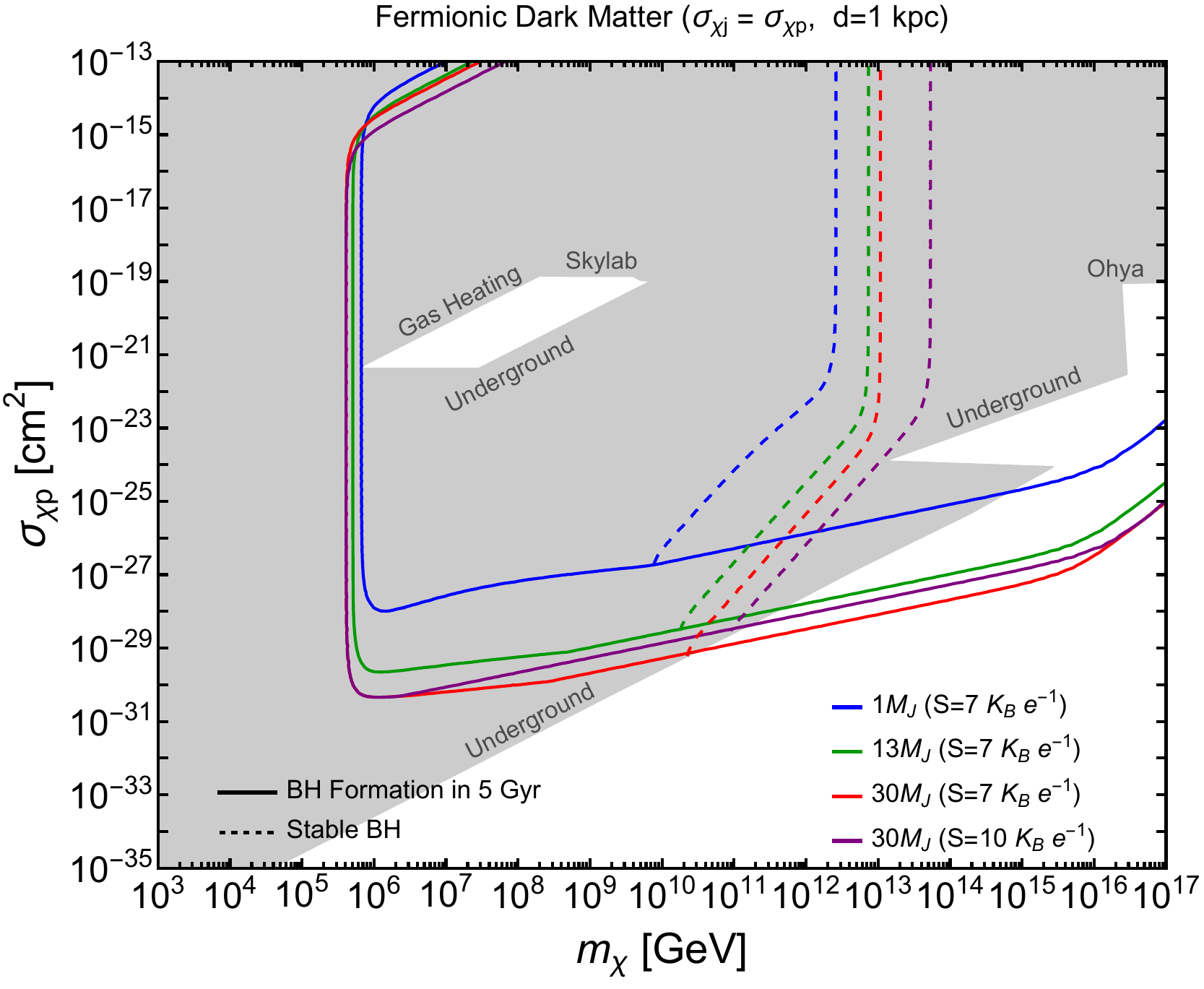}
    \caption{The parameter space where the total time for DM capture and BH formation in an exoplanet is less than $5\, \mathrm{Gyr}$, shown for different exoplanet model parameters. Solid curves delineate the boundaries of BH formation regions under various exoplanet assumptions with a distance $d=1~{\rm kpc}$ from GC within a fermionic DM model. The blue, green, and red curves correspond to exoplanet masses of $1~M_{\rm J}$, $13~M_{\rm J}$, and $30~M_{\rm J}$, respectively, with the entropy value set to $S = 7 \, k_\mathrm{B} \, {\rm e}^{-1}$. The purple curve represents an exoplanet with a mass of $30~M_{\rm J}$ and an entropy value of $S = 10 \, k_\mathrm{B} \, {\rm e}^{-1}$. Dashed curves indicate the scenario where BH mass evolution is static. Existing constraints from heavy DM searches are indicated by the gray shaded regions.
 } \label{F:CrossSectionBoundsFermionicScaledNotScaled}
\end{figure*}

The DM capture and BH formation processes depend on the temperature and density profiles of the exoplanet, which are determined by its mass and entropy parameters within the exoplanet model adopted in this study. Fig.~\ref{F:CrossSectionBoundsFermionicScaledNotScaled} illustrates the impact of different exoplanet parameters. For the entropy $S=7 \, \mathrm{k_B \, e^{-1}}$ that reproduces the properties of Jupiter, we show three benchmark exoplanet masses $1M_\mathrm{J}$ (blue), $13 M_\mathrm{J}$ (green), and $30 M_\mathrm{J}$ (red). We also show the $30 M_\mathrm{J}$ mass with an increased entropy $S=10 \, \mathrm{k_B \, e^{-1}}$ (purple).
All exoplanets are assumed to be located at $d=1\, \mathrm{kpc}$, with DM considered to be fermionic, featuring two types of scattering cross sections shown in the two panels. As the exoplanet mass increases, the capture rate becomes larger, leading to a broader DM parameter space, as observed in the comparison of the blue, green, and red curves. Similarly, since BH evolution depends on the DM capture rate, the dashed curves representing static mass evolution shift to the right with increasing exoplanet mass. The slight deviation of the upper edge arises from the increased drift time in more massive exoplanets. On the other hand, the effect of exoplanet entropy is reflected in the temperature profile. As temperature increases with higher entropy values, the thermalization time decreases, while the timescales for other conditions required for BH formation increase. As a result, the overall timescale for BH formation increases, leading to a contraction of the parameter space.

The observational signatures of superheavy DM interactions with exoplanets include the formation of BHs from planetary-mass objects and the detection of Hawking radiation from evaporating BHs. The former occurs in the parameter regions to the left of the dashed curves in Fig.~\ref{F:CrossSectionBoundsScaledNotScaled} and Fig.~\ref{F:CrossSectionBoundsFermionicScaledNotScaled}, where the surrounding planetary medium is accreted, increasing the BH mass and ultimately transforming the gas giant into a BH of approximately the same mass. On the other hand, BH evaporation takes place in the parameter regions to the right of the dashed curves. The final states of Hawking radiation may either escape the exoplanet, contributing to high-energy cosmic rays, or lose energy to the planetary medium, leading to an observable increase in its surface temperature.

The detection of planetary-mass BHs greatly benefits from existing exoplanet detection methods. In the case of exoplanets with host stars, their transmutation into BHs can still influence the motion of the host stars and be detected using Doppler spectroscopy and astrometric methods. However, since the geometric size of a BH is too small to produce observable effects in the star’s light curve, the number of transit events will decrease if exoplanets transition into BHs due to DM capture. The impact of superheavy DM can be assessed by comparing the number of observed transit events with the number of detections through Doppler spectroscopy or astrometric methods, similar to the approach proposed in~\cite{Bai:2023mfi,Bhalla:2024jbu} for detecting primordial black holes (PBHs) and dark exoplanets in extrasolar systems. While a single event cannot distinguish DM-induced BH formation from other scenarios, the large future exoplanet database may help identify it through population-level mass distributions. A notable number of missing transit events is expected in systems identified through stellar motion observations, particularly for higher-mass planets, where BH formation is more efficient. Furthermore, this reduction is anticipated to be more significant in systems closer to GC, where the DM energy density is higher.

The microlensing signals, however, appear similar for both exoplanets and BHs of the same mass due to the universality of gravitation in the point-lens limit. Since microlensing is less sensitive to the density profiles of detected objects, follow-up analysis of light curve features is essential to distinguish BHs from exoplanets of similar mass. Future improvements can be incorporated into the lensing event analyses in this direction. In \cite{Kaczmarek:2024dmp}, the distribution of lens candidates is analyzed in the two-dimensional parameter space of the Einstein timescale and microlensing parallax, aiming to distinguish BHs from other compact objects at stellar masses. However, differentiating between other classes of lenses remains challenging with current techniques. The upcoming Roman telescope will be capable of measuring the astrometric information of lenses, providing the possibility of a three-dimensional parameter space to distinguish transmuted BHs from original rogue exoplanets. In addition to identification at the single-event level, the mass distribution of planetary lenses, such as the galactic population of rogue planets recently studied in~\cite{coleman2025predictinggalacticpopulationfreefloating}, will help establish the expected event rate at a population level, providing a basis for comparison with future observations.

Gravitational microlensing is a powerful tool for detecting planet-sized objects in the Milky Way. While long-period and free-floating Jupiter-mass exoplanets have been observed through infrared surveys \cite{2000Sci...290..103Z, 2013ApJ...777L..20L, 2013Sci...341.1492D}, only microlensing techniques currently allow for the detection of Earth- to Mars-mass planets that are on wide orbits or are free-floating. This is because microlensing does not rely on the luminosity of the object of interest, but instead depends on the architecture of the planetary system, the masses of the objects in the system, and the geometry relative to a background bright lens object.

Microlensing surveys toward the Galactic bulge include the Optical Gravitational Lensing Experiment (OGLE) \cite{Szymanski:2011vx, 2015AcA....65....1U}, Microlensing Observations in Astrophysics (MOA) \cite{2001MNRAS.327..868B, 2003ApJ...591..204S}, the WISE Microlensing Survey \cite{2016MNRAS.457.4089S}, and the Korea Microlensing Telescope Network (KMTNet) \cite{2016JKAS...49...37K}. Beyond the Galactic bulge, surveys have also targeted the Magellanic Clouds, including MACHO/EROS \cite{1997A&A...324L..69R, 1998A&A...329..522R, 2001ApJ...550L.169A, 2007A&A...469..387T, 2011MNRAS.416.2949W} and the OGLE-III and OGLE-IV phases \cite{2024ApJS..273....4M, 2024ApJ...976L..19M}. In addition, microlensing observations of the Andromeda galaxy have been conducted using the Subaru Hyper Suprime-Cam (HSC) \cite{2019PhRvD..99h3503N, 2019NatAs...3..524N, 2024ApJS..273....9G}. 

Microlensing surveys have been utilized to study the abundance of PBHs and to constrain the fraction ($f$) of dark matter composed of PBHs. In a recent study, Ref. \cite{2024ApJ...976L..19M} summarized existing microlensing bounds and extended them using data from the OGLE High-Cadence Survey of the Magellanic Clouds. Additionally, Ref. \cite{2022arXiv220908215D} compiled more comprehensive constraints beyond microlensing, including those from the gamma-ray background, supernova lensing, dynamical heating of dwarf galaxies, wide binary stars, X-ray binaries, CMB distortions from accreting plasma in early universe, and disk stability arguments. This reference also presented projected future constraints from upcoming experiments such as the Roman Space Telescope, the Rubin Observatory Legacy Survey of Space and Time (LSST), and a future MeV gamma-ray facility. Among existing bounds, the strongest constraint on $f$ for Jupiter-mass black holes is $f<10^{-3}$, while future projections suggest sensitivity down to $f<10^{-4}$. 

In our work, Jupiter-sized exoplanets within the viable parameter space can transmute into black holes with masses approximately equal to that of Jupiter. To estimate the total mass of such transmuted black holes, we assume there is, at most, roughly one Jupiter-sized planet per star in the Milky Way (see Ref. \cite{2017Natur.548..183M} for estimates of the free-floating exoplanet population). Given that the Milky Way contains on the order of $10^{11}$ stars, the total mass of Jupiter-sized objects would be approximately $10^8 \, M_{\odot}$. If these exoplanets were to fully transmute into black holes, and assuming the dark matter halo mass is on the order of $10^{12} \, M_{\odot}$, the resulting maximum mass fraction from this mechanism alone would be $ f \sim 10^{-4}$, which is comparable to current and projected upper limits from microlensing constraints. Therefore, our scenario of Jupiter-mass black hole formation from exoplanet collapse is not excluded by existing bounds on PBHs. Future surveys will be particularly valuable, as they will probe with greater sensitivity and place stronger constraints on $f$. This will help further test the viability of the exoplanet-collapse scenario as a possible origin of Jupiter-mass black holes.

The initial mass of BHs formed from the capture of heavier DM masses, as indicated by the region for evaporating BHs, is very small, causing them to evaporate into all particle species with masses smaller than the Hawking temperature. The Hawking radiation particles are produced at the center of the exoplanet, where BH formation occurred. The signal channel for detecting BH evaporation inside exoplanets is contingent on the propagation of high-energy particles through the exoplanet, which in turn depends on its interior profile and the Hawking radiation energy spectrum, determined by the BH mass. For particles with a mean free path larger than the exoplanet’s radius, Hawking radiation particles can contribute to high-energy cosmic rays, potentially including photons, electrons, positrons, and neutrinos, serving as targets for indirect detection observations. Similarly, DM particles can also be produced via Hawking radiation if the Hawking temperature is sufficiently large, leading to a novel flux of boosted DM that can be detected in direct detection experiments. On the other hand, Hawking radiation particles contribute to the heating of exoplanets if their mean free path is smaller than the radius of the exoplanets. The heating effect can be detected by measuring thermal emissions from the surface of exoplanets using infrared and optical telescopes.

\begin{figure}[t]
    \centering
    \includegraphics[width=0.5\columnwidth]{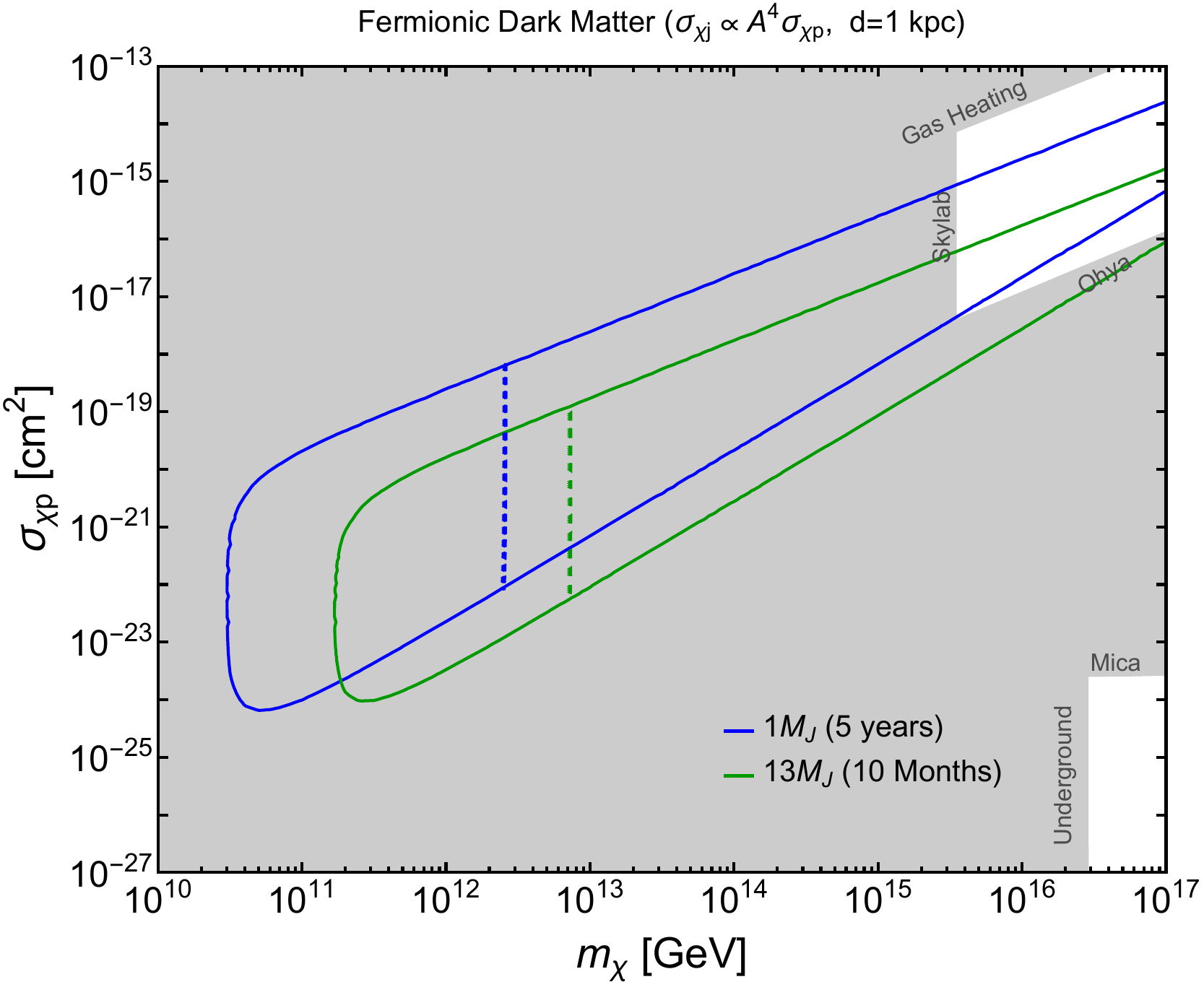}
    \caption{Parameter space for periodic BH formation. The DM is assumed to be a fermion with spin-independent scattering. BH formation occurs within $5\, \mathrm{years}$ for a planet with a mass of $1M_\mathrm{J}$ (blue) located at $d=1\,\mathrm{kpc}$ from GC, and within $10\, \mathrm{months}$ for a planet with a mass of $13 M_\mathrm{J}$ (green) at the same location. Existing constraints are indicated by the gray shaded regions. } \label{F:PeriodicFormation}
\end{figure}

Another notable feature of DM-induced BH formation is that sustained DM capture enables the process to occur periodically, particularly in regions with high DM mass and large cross section. Thus observation signals of pulsed high-energy cosmic rays and periodic variations in exoplanet temperature are also expected to manifest on the timescale of BH formation. In Fig.~\ref{F:PeriodicFormation}, we present the parameter space for BH formation on short time scales. The blue curve represents BH formation within $5~{\rm years}$ for a planet with a mass of $1 M_{\rm J}$, while the green curve corresponds to formation within $10~{\rm months}$ for a planet with a mass of $13 M_{\rm J}$. The planet is assumed to be located at $d=1\, {\rm kpc}$, and the DM is assumed to be a fermion with a spin-independent cross section. In this case, the slope of the lower edge is primarily determined by the thermalization time rather than the capture time. Compared to the existing constraints shown in the shaded gray region, we observe that periodic BH formation occurs within DM parameter regions that remain viable. BHs formed in this region have small initial masses and will evaporate into energetic particles, as discussed above. The flux of the Hawking evaporation signal observed at Earth is suppressed by the distance to the source; therefore, stronger signals are expected from nearby exoplanets, where the lower edge of the periodic BH formation region will be slightly higher (see the discussion of Fig.~\ref{F:CrossSectionBoundsScaledNotScaled} for details).

Finally, we discuss the high-energy particle signals from BH formation and evaporation. To illustrate this, we first consider BH formation at three benchmark points within the blue contour, corresponding to BHs forming within $5~{\rm years}$ in an exoplanet of mass $1M_\mathrm{J}$ located at $1 \, \mathrm{kpc}$ from GC. For a DM mass and cross section of $m_\chi = 3 \times 10^{12} \, \mathrm{GeV}$ and $\sigma_{\chi {\rm p}} = 10^{-21} \, \mathrm{cm^2}$, which is close to the static mass evolution indicated by the dashed curve, the initial BH mass is estimated to be $M_\mathrm{BH}^\mathrm{init} \simeq 7.4 \times 10^9 \, \mathrm{g}$. The corresponding BH Hawking temperature, given by $T_\mathrm{BH} = 1 / 8 \pi G M_\mathrm{BH}$, is approximately $1.4 \, \mathrm{TeV}$, representing the energy of the evaporation final states. The BH evaporation time, calculated using Eq.~\eqref{E:BlackHoleTime}, is approximately one day. Notably, the lifetime is longer than in the case of pure Hawking radiation, as the subdominant mass accretion delays the final evaporation. As another example, for $m_\chi = 4 \times 10^{15} \, \mathrm{GeV}$ and $\sigma_{\chi {\rm p}} = 10^{-17} \, \mathrm{cm^2}$, the initial mass, evaporation time, and temperature of the BH are $M_\mathrm{BH}^\mathrm{init} \simeq 1.5\times 10^5 \, \mathrm{g}$, $t_\mathrm{BH} \simeq 3 \times 10^{-10} \, \mathrm{s}$, and $T_\mathrm{BH} \simeq 7 \times 10^7 \, \mathrm{GeV}$, respectively. Similarly, for $m_\chi = 10^{17} \, \mathrm{GeV}$ and $\sigma_{\chi {\rm p}} = 10^{-15} \, \mathrm{cm^2}$, the corresponding values are $M_\mathrm{BH}^\mathrm{init} \simeq 1.2 \times 10^3\, \mathrm{g}$, $t_\mathrm{BH} \simeq 1.6 \times 10^{-16} \, \mathrm{s}$, and $T_\mathrm{BH} \simeq 9 \times 10^{9} \, \mathrm{GeV}$. The last two benchmarks correspond to the allowed DM parameter space delineated by the gas heating, Skylab, and Ohya constraints. Periodic BH formation offers an additional avenue for probing this parameter space.

The propagation of Hawking radiation final states depends on their energy at the time of production, which, as discussed earlier, reaches very high energies for the parameter space of interest. To assess whether such high-energy photons can escape the exoplanet or become trapped within it, we estimate their mean free path based on studies of energetic cosmic ray propagation in \cite{Lee:1996fp}. At high energies, the triplet pair production (TPP) process, $\gamma e^{-} \to e^{-} e^{+} e^{-}$, dominates the interaction rate, and the cross section saturates with a mild logarithmic dependence on the collision energy at $\sigma_{\rm TPP}\sim\mathcal{O}(10^{-33}) \, \mathrm{cm^2}$. Given the range of the TPP cross section and the assumption that the electron number density equals the proton number density, we find that photons undergo multiple scatterings within the interior medium before reaching the planet’s surface. Energetic leptons are produced both through the TPP process and directly from Hawking radiation, and their propagation in the hydrogen-helium medium requires a more comprehensive analysis. Additionally, the presence of the exoplanet’s magnetic fields further complicates the propagation of charged particles. We leave a detailed study of the propagation model in exoplanets and the observation methods for future work.


\section{Conclusion}
\label{S:Conclusion}
We conducted a complementary study to constrain non-annihilating superheavy DM models using exoplanets as a probe. DM can be captured by exoplanets, accumulate in their interiors, and potentially form a BH at their center. The constraints in this study are based on the assumption that a single BH can form within the lifetime of exoplanets. The existence of exoplanets places bounds on the scattering cross section and mass of DM, as BHs can grow and convert exoplanets into subsolar-sized BHs. Additionally, the formed BH at larger DM masses is light enough to evaporate and vanish due to Hawking radiation, which predominantly emits high-energy particles.

We adopted a simplified model for the structure of exoplanets and demonstrated that their diversity in mass and location enables them to probe a broader DM parameter space not yet excluded by terrestrial experiments. We tested the sensitivity of the bounds to the parameters chosen for exoplanet modeling, including variations in their interior density and temperature profiles, as well as the location and mass of the exoplanets. Additionally, we studied four types of DM models: fermionic vs. bosonic, and with vs. without a coherent effect in the scattering cross section. We showed that, in parameter spaces not excluded by existing bounds, it remains possible for a BH to form within time scales as short as ten months. This implies that, over the lifetime of an exoplanet, a BH could form multiple times.

This work can be extended in four directions in future studies. Firstly, we did not consider self-scattering in this analysis. Since self-scattering is highly effective in kinetic energy loss, we expect changes in the bounds for larger DM masses, where scattering with baryonic matter becomes inefficient. Secondly, smaller BHs in the upper range of the bounds evaporate into highly energetic particles. While photons are more likely to decay within exoplanets, the resulting electron-positron pairs may escape, and investigating the phenomenology of such radiation requires a dedicated study. Thirdly, the multiple occurrences of BH formation within exoplanets could have intriguing consequences. For example, if each BH remains stable, exoplanets may continuously radiate particles. Lastly, the exoplanetary probe can be refined with improved modeling of the internal structure of planets. For instance, the temperature profile alters DM drifting and BH formation time scales, while the density profile determines the capture rates. Additionally, changes in the temperature profile affect the fusion processes that shape the planet’s chemical composition, potentially influencing the spin-independent scattering rate.


\begin{acknowledgments}
The authors would like to thank Tao Xu and Hai-Bo Yu for their encouragement, helpful discussions, and insightful comments on the manuscript. We would like to thank Joseph Bramante for helpful discussions. M.P.-M. is supported by the U.S. Department of Energy under Grant No. DE-SC0008541. T.F. acknowledges support from an appointment through the NASA Postdoctoral Program at the NASA Astrobiology Center, administered by Oak Ridge Associated Universities under contract with NASA.
\end{acknowledgments}


\appendix


\section{Dark Matter Velocity Distribution}
\label{A:Velocities}
In this appendix, we explain the details of the DM halo parameters and the model for the distribution of luminous matter used in our velocity calculations. Following \cite{Borukhovetskaya:2021ahz}, we adopt the NFW density profile for the Milky Way DM halo \cite{Navarro:1996gj},
\begin{equation}
   \rho_\mathrm{\chi}\left(r\right)=\frac{\rho_{\rm s}}{(r/r_{\rm s})\left(1+r/r_{\rm s}\right)^2},
\end{equation}
with the scale radius $r_{\rm s}=20.2\, \mathrm{kpc}$, the characteristic density $\rho_{\rm s}=0.29 \, {\rm GeV \, cm^{-3}}$, the virial radius of $R_{\rm 200}=191.9\, \mathrm{kpc}$, and the total mass up to $R_{200}$ equal to $1.15\times 10^{12} \, {\rm M_{\odot}}$. The gravitational potential of the NFW profile is
\begin{equation}
   V_\mathrm{\chi}(r)=-4\pi G r_{\rm s}^2 \rho_{\rm s} \frac{\ln{\left(1+r/r_{\rm s}\right)}}{r/r_{\rm s}}. 
\end{equation}
We assume the central baryonic matter distribution in the Milky Way galaxy consists of a bulge, a thick disk, and a thin disk. For the bulge, we assume the Hernquist model \cite{Hernquist:1990be},  
\begin{equation}
   \rho_\mathrm{B}(r)=\frac{M_\mathrm{B}}{2 \pi} \frac{a_\mathrm{B}}{r\left(r+a_\mathrm{B}\right)^3},
\end{equation}
with the total mass of $M_\mathrm{B}=2.1\times10^{10} \, {\rm M_{\odot}}$ and $a_\mathrm{B}=1.3\, \mathrm{kpc}$. The potential associated with the Hernquist profile is
\begin{equation}
   V_\mathrm{B}(r)=-\frac{G M_\mathrm{B}}{\left(r+a_\mathrm{B}\right)}. 
\end{equation}
For the thick disk and thin disk components of the Milky Way, we adopt the Miyamoto-Nagai model \cite{Miyamoto:1975zz}, where the density in cylindrical coordinates, defined by the radial distance $r$ and height $z$, is 
\begin{align}
\rho_{\rm D}\left(r,z\right)=\frac{b_\mathrm{D}^2 M_\mathrm{D}}{4 \pi}
 \frac{a_\mathrm{D} r^2+\left( a_\mathrm{D}+3\sqrt{z^2+b_\mathrm{D}^2}\right)\left(a_\mathrm{D}+\sqrt{z^2+b_\mathrm{D}^2}\right)^2 }{\left(r^2+\left(a_\mathrm{D}+\sqrt{z^2+b_\mathrm{D}^2}\right)^2\right)^{\frac{5}{2}}\left(z^2+b_\mathrm{D}^2\right)^{\frac{3}{2}}}, 
\end{align}
and the corresponding gravitational potential is 
\begin{equation}
   V_{\rm D}\left(r,z\right)=- \frac{G M_\mathrm{D}}{\left(r^2+\left(a_\mathrm{D}+\sqrt{z^2+b_\mathrm{D}^2}\right)^2\right)^{\frac{1}{2}}}.
\end{equation}
For the thick disk, the total mass is $M_\mathrm{D}=2.0\times10^{10} \, {\rm M_{\odot}}$, $a_\mathrm{D}=4.4\, \mathrm{kpc}$, and $b_\mathrm{D}=0.92\, \mathrm{kpc}$. For the thin disk, $M_\mathrm{D}=5.9\times10^{10} \, {\rm M_{\odot}}$, $a_\mathrm{D}=3.9\, \mathrm{kpc}$, and $b_\mathrm{D}=0.31\, \mathrm{kpc}$. In the end, the total gravitational potential of the Milky Way galaxy is given by
\begin{equation}
   V_{\rm MW}\left(r, z\right)=V_\mathrm{\chi}+V_{\rm B}+V_{\rm D}^{\rm Thick}+V_{\rm D}^{\rm Thin}.
\end{equation}
Using the adopted potential, we determine the velocity distribution of DM particles, with their circular velocity $v_\mathrm{r}$, 1D velocity dispersion $\sigma_{\rm v}$, and escape velocity $v_\mathrm{eh}$, as functions of the distance from the GC\footnote{Note that the DM velocity distributions is calculated assuming $z=0$ for exoplanets located on the plane of the galactic disk. We checked that for $z\leq 2~{\rm kpc}$, the difference remains within $6\%$ for $v_\mathrm{r}$, $0.9\%$ for $v_\mathrm{eh}$, and $3\%$ for $\sigma_{\rm v}$, at $r=8~{\rm kpc}$. At a distance of $r=1~{\rm kpc}$, the difference is below $4.5\%$ for $v_\mathrm{r}$, $3.8\%$ for $v_\mathrm{eh}$, and $8.3\%$ for $\sigma_{\rm v}$.}
\begin{eqnarray}
v_\mathrm{r}(r)^2 &=& r \frac{dV_{\rm MW}\left(r,0\right)}{dr}, \\
v_\mathrm{eh}(r)^2 &=& -2 V_{\rm MW}\left(r, 0\right), \\
\sigma_{\rm v}(r)^2 &=& \frac{1}{\rho_\mathrm{\chi}\left(r\right)}\int_r^{\infty} \rho_\mathrm{\chi}\left(r'\right)\frac{dV_{\rm MW}\left(r',0\right)}{dr'} dr'. 
\end{eqnarray}

\begin{figure}[t]
    \centering
    \includegraphics[width=0.45\columnwidth]{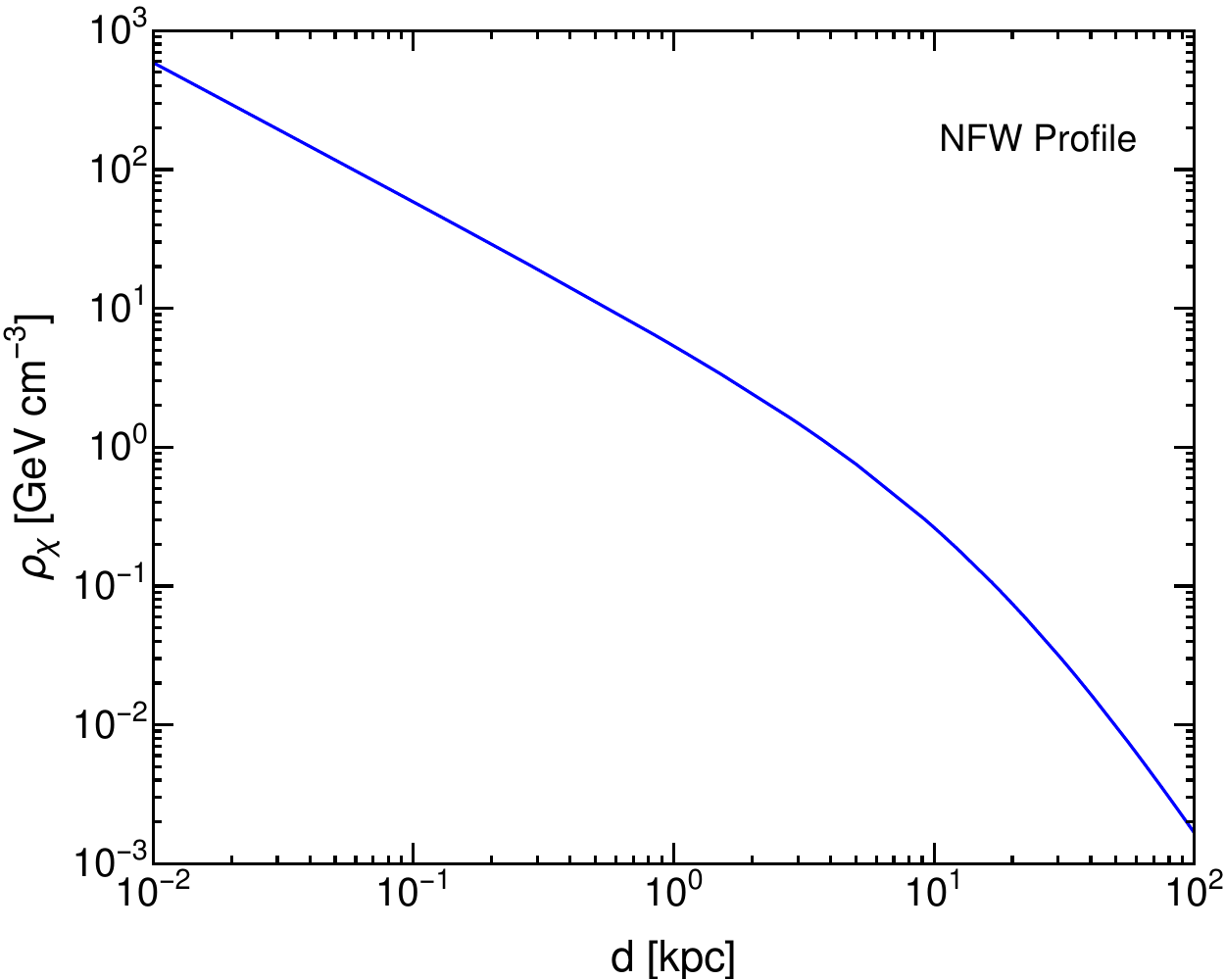}\,\,\,
    \includegraphics[width=0.4435\columnwidth]{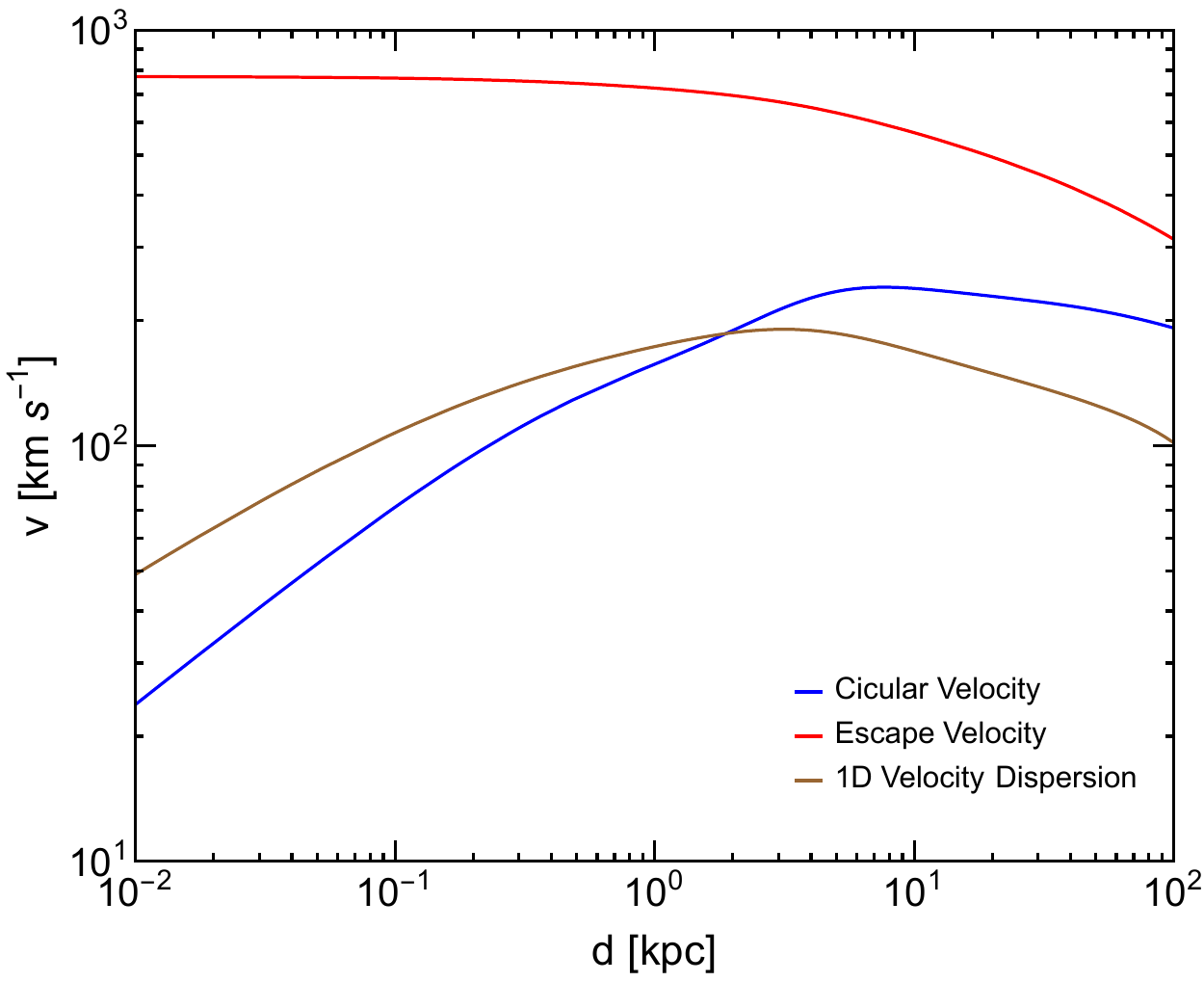}
    \caption{Left: DM density profile of the Milky Way halo. Right: Velocity profiles for the circular velocity (blue), escape velocity (red), and 1D velocity dispersion (brown). The DM component follows the NFW profile, and the luminous component includes a Hernquist model for the bulge and Miyamoto-Nagai models for the thick and thin disks.} 
    \label{F:Velocities_NFW}
\end{figure}

Fig.~\ref{F:Velocities_NFW} depicts the Milky Way DM halo profile adopted for this work. The left panel shows the density profile. The right panel displays the velocities, including the circular velocity (blue), galactic escape velocity (red), and 1D velocity dispersion (brown).

\bibliography{references}

\end{document}